\documentclass[aps,10pt,twocolumn,superscriptaddress]{revtex4}
\usepackage{graphicx}
\usepackage{amssymb, amsmath}
\usepackage{color}
\usepackage{ulem}
\usepackage{graphicx}
\usepackage{dcolumn}
\usepackage[caption=false]{subfig}
\def\I{\uppercase\expandafter{\romannumeral 1}}
\def\II{\uppercase\expandafter{\romannumeral 2}}
\def\III{{\uppercase\expandafter{\romannumeral 3}}}
\def\IV{{\uppercase\expandafter{\romannumeral 4}}}
\def\V{{\uppercase\expandafter{\romannumeral 5}}}
\def\VI{{\uppercase\expandafter{\romannumeral 6}}}
\def\VII{{\uppercase\expandafter{\romannumeral 7}}}
\def\i{\lowercase\expandafter{\romannumeral 1}}
\def\ii{\lowercase\expandafter{\romannumeral 2}}
\def\iii{{\lowercase\expandafter{\romannumeral 3}}}
\def\iv{{\lowercase\expandafter{\romannumeral 4}}}
\def\v{{\lowercase\expandafter{\romannumeral 5}}}
\def\vi{{\lowercase\expandafter{\romannumeral 6}}}
\def\vii{{\lowercase\expandafter{\romannumeral 7}}}

\def\nn{\nonumber\\}

\def\nn{\nonumber\\}

\def\s{\textrm{s}}
\def\t{\textrm{t}}

\begin{document}

\title{Floquet spin and spin-orbital Hamiltonians and doublon-holon generations in periodically driven Mott insulators}

\author{Kasra Hejazi}
\affiliation{ Department of Physics, University of California, Santa Barbara
CA 93106, USA}

\author{Jianpeng Liu}
\affiliation{ Kavli Institute for Theoretical Physics, University of California, Santa Barbara
CA 93106, USA}

\author{Leon Balents}
\affiliation{ Kavli Institute for Theoretical Physics, University of California, Santa Barbara
CA 93106, USA}

\begin{abstract}
We consider Mott insulators driven by periodic coherent laser radiation, using both single orbital and multi-orbital models, noting that the latter is of more interest in solid state systems. We derive general expressions for the resulting periodically driven spin models and spin-orbital models using time-dependent perturbation theory. First, we show that the effective exchange interactions of the Floquet Hamiltonians are highly tunable by the frequency, amplitude, and polarization of the laser. Second, we take the effect of finite bandwidth of excitations into account and study possible heating effects. Using the same formalism with a slight modification we also consider the small frequency regime and study the dielectric breakdown of Mott insulators.
\end{abstract}

\maketitle

%- Opening:

The study of periodically driven quantum systems has received significant attention in recent years.
A common theoretical prescription is the Floquet formalism \cite{floquet-pr65, floquet-pra73}, which amounts to finding the eigenstates of the time evolution operator $U(T+t , t)$ from time $t$ to $t+T$, where $T$ is the period of the drive.  These states have the form $e^{- i \epsilon_n t} \vert n \rangle_t$, with $\vert n \rangle_t$ a periodic state  with the same period as that of the drive and $\epsilon_n$ called the quasi-energy. This form for eigenstates ultimately allows for a description of a time-periodic system using some time-independent Hamiltonian dubbed as the ``Floquet Hamiltonian", 
$H_{\textrm{F}}=i\hbar\log{U(T,0)}/T$, where $U(T,0)$ is the stroboscopic time-evolution operator from time $0$ to a full period $T$.   One can further write down the evolution operator from arbitrary time $t_0$ to another arbitrary time $t$ with the use of the operators called the micromotion operators, as $U(t,t_0) = \hat{U}_F(t) \, e^{- i H^{\text{eff}} (t - t_0)} \, \hat{U}_F^\dagger (t_0)$, where $H^{\text{eff}}$ is a time-independent \textit{effective} Hamiltonian and $\hat{U}_F(t)$, the micromotion operator, is a periodic operator yielding intraperiod dynamics \cite{eckardt2015high,bukov-aip-15}.

%- literature review on Floquet stuff

A natural way to periodically drive a condensed matter system is with electromagnetic radiation. Since the details of the Floquet Hamiltonian describing this situation are crucially dependent on the frequency, amplitude and polarization of the external drive, one is able to engineer the physical properties of a quantum system to a large extent using laser-light radiation\cite{basov2017towards}. Such ``Floquet engineering" has been extensively studied in the context of both single-particle and many-body condensed-matter physics. In non-interacting systems, the light radiation dresses the electronic band structure, which may change the topological character, leading to  various exotic phenomena\cite{takashi-aoki-prb09, floquet-transport-prb11,floquet-ti-np11, tanaka-prl10, floquet-silicene-prl13, platero-prl13, floquet-classification-prb11, levin-prx13,lindner-prx16, floquet-ti-surface-exp, floquet-volkov-np16}. On the other hand, in interacting systems the Floquet physics has been explored in the context of the light-induced/light-enhanced superconducting \cite{floquet-cuprate-science11, floquet-cuprate-nature14, floquet-k3c60-nature16, floquet-pairing-prb16, floquet-pairing-arxiv17}, charge-density-wave \cite{floquet-cdw-prl13}, and spin-density-wave \cite{floquet-sdw-nm12} Fermi-surface instabilities.  Laser controlled exchange interactions in single-band Mott insulators \cite{floquet-exchange-nc15}, topological phase transitions in Kondo insulators \cite{floquet-kondo-arxiv17}, and the possible periodically driven topologically ordered states \cite{floquet-chiral-sl-arxiv16, floquet-fractional-chern-insulator-prl14} have also been discussed.

%- Exponentially slow heating

In the presence of interactions, one expects a periodically driven system to eventually heat up to infinite temperature at long times \cite{floquet-infiniteT-prx14, moessner-pre14}, i.e.~ that the density operator of any finite subsystem become maximally mixed and featureless at long times. However, as has been shown rigorously, the heating rate can be (quasi-)exponentially slow \cite{slow-heating-prl15,abanin2017rigorous} in the ratio of the driving frequency to the local energy scales.  As a result of this, one expects such system to show interesting prethermal behavior. One can even think of the possibility for a system to first relax into a steady prethermal state at intermediate times $t\lessapprox\tau^{*}$, then ultimately evolve into the infinite temperature state at long times $t\gtrapprox\tau^{*}$. Such a prethermal regime is realized numerically in \cite{machado2017exponentially}, where it has been shown in a lattice spin model that such a system can first equilibrate to a (pre)thermal state with respect to a time-independent effective Hamiltonian before it reaches the infinite temperature state. Such time-independent effective Hamiltonians are dubbed as ``Floquet Hamiltonians" in the literature, and are typically expressed as series expansions in $1/\omega$.  Furthermore, it has been explicitly shown in \cite{machado2017exponentially} using numerics, that the characteristic infinite temperature time scales $\tau^{*}$ in their models grow exponentially with the increase of the driving frequency.

%- multi-orbital physics
We will study periodically driven Mott insulators in this work. We start with single-orbital Mott insulators, i.e. the Hubbard model, at half filling. The effective spin Floquet Hamiltonian for periodically driven half-filled Hubbard model has been derived \cite{floquet-exchange-nc15,floquet-chiral-sl-arxiv16,sw-bukov-prl16}. In the static case, in an insulator, the Hubbard model has two relevant energy scales: (\i) the energy scale of spin dynamics, i.e.~exchange interaction $J$ between the electrons at neighboring sites, and (\ii) the onsite electron-electron interaction energy $U$ which comes into play when there are doubly occupied sites. In the Mott insulating regime, the latter is much larger than the former, $U \gg J$. The periodically driven Hubbard model, on the other hand, has another energy scale that is the driving frequency $\omega$.   Following similar arguments as in the previous paragraph, one should be concerned about the regime in which a rapid heating does not occur in this system. There are two classes of processes that can lead to heating of the system due to the absorption of photons: one is by multi-spin re-orderings, and the other is by creation of doubly-occupied sites in the system. Considering the second case, one can think of the doubly occupied sites (doublons) and the empty sites (holons) which are created as a consequence as new dynamical degrees of freedom. Due to the hopping of electrons in the original Hubbard model, the doublons and the holons are able to hop around and thus these excitations of the system have a non-vanishing bandwidth \cite{brinkman-rice-prb70,kane-lee-read}. If photons that strike the system are able to supply an energy that lies within this bandwidth, one expects to see a rapid heating due to creation of doublon-holon (DH) pairs in the system. As we will see, heating can be avoided if the frequency is kept outside of certain resonant windows, and at the same time also kept much larger than the effective spin exchange. In this paper, we will restrict our attention to states with very low density of DH pairs, and  will develop a time dependent perturbation theory that will take the above points into account.

Most of the previous theoretical studies of similar Floquet systems have been focused on spin degrees of freedom and the electron-phonon couplings. To the best of our knowledge, the orbital degrees of freedom and their interplay with the spins have never been addressed in the context of Floquet physics. This is most relevant to solid-state Mott insulators like titanates, nickelates, and manganites. Given that the orbitals play essential roles in strongly correlated transition-metal oxides \cite{kugel1982jahn, tokura-orbitalreview, georges-review13}, in this paper we next consider driving multi-orbital Mott insulators using laser radiation. We use multi-orbital Hubbard models to describe such systems, with the on-site electron interactions much greater than the hopping parameters, and we consider the filling equal to one electron per site. In a multi-orbital Mott insulator, the on-site interaction energy depends on the spin and orbital configurations of the electrons at a multiply occupied site. Furthermore, for a hopping event between two given sites, the hopping parameters can also depend on the initial and final orbital configurations of the sites. This added complexity of multi-orbital Mott insulators has an upside: it introduces more freedom to engineer the exchange interactions in the effective Floquet Hamitlonian.

Based on time-dependent perturbation theory, we first derive general expressions for the time evolution in the periodically driven spin and spin-orbital models. Including the effects of the DH hoppings, i.e.~taking the effect of the bandwidth of excitations into account in our perturbation theory, we find that the Floquet Hamiltonian projected onto a generic state in the zero-doublon subspace contains both real and imaginary parts. The real part is interpreted as an effective spin or spin-orbital model, and the corresponding exchange interactions are renormalized by the periodic driving, which allows for the Floquet engineering of the interactions. The imaginary part on the other hand is related to the rate of generation of DH pairs, and thus can capture the effects of heating, due to the increase in the density of DH pairs. With all this said, one can work in two different regimes using the formalism of this paper. Either one is away from a resonance and not too many DH pairs are created and thus a spin(-orbital) effective Hamiltonian captures the physics well, or one is inside one of the resonant windows and the physics of the system, at least for short time, is described by studying how DH pairs density increases as a result of resonant radiation. We furthermore study, by slightly altering the formalism, the creation rate of DH pairs at very small frequencies, and show that indeed a non-trivial zero-frequency limit exists. The results in this limit can be interpreted as the behavior of the system when exposed to static external field, and thus is a reflection of the (static) field-induced breakdown of a Mott insulator.

In a prior short paper\cite{liu2018floquet}, some parts of this formalism were presented, and applied to the orthorhombic titanates YTiO$_3$ and LaTiO$_3$ using first-principles calculations. It was observed that as a result of multi-orbital interactions,  ferromagnetic and antiferromagnetic Mott insulators exhibit distinct responses to laser radiation.  The effective exchange interactions in these titanates may be engineered to a large extent, and may be even flipped at moderate electric-field energies.  The present paper derives and extends the formalism of this earlier work, and discusses in much more detail the physics of doublon generation. 

 The remainder of this paper is organized as follows. In Section \ref{sec:floquet_spin} we discuss the formalism of the Floquet spin model derived from the periodically driven Hubbard model and apply it to single orbital Hubbard model. In Section \ref{sec:floquet_spin_orbital} we generalize the formalism to the case of multiorbital Mott insulators. We finally present a summary of what has been done in the paper.

\section{Floquet Spin Model}
\label{sec:floquet_spin}

We start the discussion with a half-filled single-orbital Hubbard model which is periodically driven by laser radiation. Such a problem has been discussed in \cite{floquet-exchange-nc15,floquet-chiral-sl-arxiv16,sw-bukov-prl16}. Here we re-derive the effective Floquet spin Hamiltonian using time-dependent perturbation theory, and show that one can capture novel physics if one takes the effect of the finite bandwidth of the excitations into account. We later will use similar methods to generalize the discussions to multi-orbital Mott insulators.

\subsection{Time Dependent Perturbation Theory}
 
We consider the following periodically driven Hubbard model:
\begin{equation}
H(t)=-\sum_{\langle ij\rangle \sigma} \left( t_{h}\,e^{i\,u_{ij}\,\sin\omega t} \ c_{i\sigma}^{\dagger} c_{j\sigma}
 + \text{h.c.} \right) + \ U\sum_{i}\hat{n}_{i\uparrow}\hat{n}_{i\downarrow}\;,
 \label{eq:hubbard}
\end{equation}
where $t_h$ is the hopping amplitude between sites $i$ and $j$, and $U \gg t_h$ is the onsite Coulomb repulsion energy. $u_{ij}\!=\!e\mathbf{E}_0\cdot\mathbf{r}_{ij}/ \omega$ (we have set $\hbar=1$),
where $\left|\mathbf{E}_0\right|$ denotes the magnitude of the oscillating electric field of a laser with frequency $\omega$, $\mathbf{E}(t)\!=\!\mathbf{E}_0\cos{\omega t}$, and $\mathbf{r}_{ij}$ is the displacement vector between two lattice sites $i$ and $j$. Only the nearest neighbor hopping is taken into account here and the model is studied at half-filling. Note that Hermiticity requires $u_{ij} = - u_{ji}$.

In equilibrium without driving it is well known that at half filling the system stays in the Mott insulating phase in the limit $U\!\gg\!t_{h}$. Then the low energy physics is dominated by the spin dynamics, and is well described by a Heisenberg model with antiferromagnetic nearest neighbor exchange $J_{ij}=4t_{h}^2/U$, this result can be derived using a second-order time independent perturbation theory \cite{fazekas-book}.  We generalize the discussions to the case with periodic laser radiation, and derive a \textit{time-dependent} spin model using time-dependent second-order perturbation theory. 

A generic many-body state $|\Psi\rangle_t$ can be expressed as a linear superposition of states with $n$ doubly occupied sites (which are dubbed as ``doublons"): $\vert\Psi\rangle_t\!=\!\sum_{n=0}^{\infty}\vert\Psi_n\rangle_t$, where $\vert\Psi_n\rangle_t$ represents the component of the state of the system with $n$ doublons, i.e., $U\sum_{i}\hat{n}_{i\uparrow}\hat{n}_{i\downarrow}\vert\Psi_n\rangle_t = n\,U\,\vert\Psi_n\rangle_t$. The Schr\"odinger equation for the evolution of the different components of the state of the system reads:
 \begin{widetext}
 \begin{equation}
	\begin{aligned}
		i \partial_t |\Psi_{0}\rangle_t = \hat{P}_0 \ H \ |\Psi \rangle_t &= \hat{P}_0 \ T_t |\Psi_1\rangle_t \ , \\
		i \partial_t |\Psi_{1}\rangle_t = \hat{P}_1 \ H \ |\Psi \rangle_t &=  U \ |\Psi_{1}\rangle_t + \ T_t |\Psi_0\rangle_t + \hat{P}_1 \ T_t |\Psi_1\rangle_t + \hat{P}_1 \ T_t |\Psi_2\rangle_t  \ ,  \\
		& \vdots
	\end{aligned}
\end{equation}
\end{widetext}
 where $T_t = -\sum_{\langle ij\rangle \sigma} \left( t_{h}\,e^{i\,u_{ij}\,\sin\omega t} \ c_{i\sigma}^{\dagger} c_{j\sigma}
 + \text{h.c.} \right) $ is the time dependent hopping term in the Hamiltonian \eqref{eq:hubbard} and $\hat{P}_n$ is the projector onto the subspace with $n$ double occupancies. 
 
Since we are interested in the dynamics of states with a small local density of doublons and holons, we approximately consider only the dynamics induced by $\vert \Psi_1 \rangle_t$ for the component $\vert  \Psi_0 \rangle_t$, and neglect corrections due to the effects of ${\vert \Psi_n \rangle_t}$, with $n>1$. We claim that the essential properties of the dynamics of the system can be captured by this approximation. Thus we will continue by neglecting the $\hat{P}_2 \vert \Psi\rangle, \hat{P}_3 \vert \Psi\rangle, \ldots $ components of the time dependent state in the above Schr\"odinger equation and focusing on how $\vert\Psi_0\rangle$ and $\vert\Psi_1\rangle$ evolve mutually.   Note that these higher order components will contribute with at least the fourth order of $t_h$ to the time evolution of the spin state. The truncated equations of motion take the following form:
 \begin{equation}\label{eq:truncated_eom}
	\begin{aligned}
		i \partial_t |\Psi_0\rangle_t &=  \hat{P}_0 \ T_t |\Psi_1\rangle_t ,\\
		i \partial_t |\Psi_{1}\rangle_t &=  U \ |\Psi_{1}\rangle_t + \ T_t |\Psi_0\rangle_t + \hat{P}_1 \ T_t |\Psi_1\rangle_t .
	\end{aligned}
\end{equation}
The hopping operator $P_1 \ T_t $ in the second line of \eqref{eq:truncated_eom} can be replaced by $\tilde{T}_t = P_1 T_t P_1$ since it is acting on the one double occupancy subspace. One can think of the action of the operator $\tilde{T}_t$ as the hopping operator of the doublon and holon restricted to the 1 doublon-holon (DH) pair subspace. Note that we have kept the term $T_t |\Psi_1\rangle_t$ in the above equation, although it will give the same order corrections to the dynamics of $\vert \Psi_0 \rangle_t$ as the terms that are neglected. The reason is that it accounts for the effects of the finite bandwidth of excitations, which can give rise to a form of heating due to the creation of doublon holon pairs. We will discuss these matters more in what follows.

We will restrict our attention for now to the regime in which the frequency is much larger than the hopping amplitude $t_h$; one continues with \eqref{eq:truncated_eom}, by approximating the hopping operator within the single DH subspace $\tilde{T}_t$ by its time average, (Appendix \ref{app:time_averaging}) 
 \begin{equation}
 	\begin{aligned}
 		\bar{T} &= \frac{\omega}{2\pi} \int_0^{2\pi/\omega} dt' \ \tilde{T}_{t'} \\
 		&= t_h \sum_{\langle i j \rangle} \left[ \left(\mathcal{J}_0(u_{ij}) \ \hat{P}_1 \, \hat{v}_{ij}  \, \hat{P}_1 \right) + \left( i \leftrightarrow j \right) \right],
 	\end{aligned}
 \end{equation}
  where $\mathcal{J}_n$ stands for the Bessel function of the first kind and $\hat{v}_{ij} = \left( - \sum_\sigma c^\dagger_{i\sigma} c_{j\sigma} \right)$ is the hopping operator for the electrons. The operator $\hat{P}_1 \, \hat{v}_{ij}  \, \hat{P}_1$ is the hopping operator for the doublon and the holon. The second equation of \eqref{eq:truncated_eom} reads thus:
 \begin{equation}
	e^{-i (U +  \bar{T}) t}  \ i \, \partial_t   \left( e^{i (U +  \bar{T}) t} \lvert \Psi_{1} \rangle_t \right)  =   T_t \lvert\Psi_{0} \rangle_t .
\end{equation}
 Integrating both sides, and fixing the initial conditions such that the lower limits of the integrals cancel each other, one arrives at:
 \begin{equation}
	 \begin{aligned}
 			i\ &  \left( e^{i (U +  \bar{T}) t} \lvert \Psi_{1} \rangle_t \right)  =  t_h \int^t dt' \; e^{i (U +  \bar{T}) t'} \ \sum_{\langle ij\rangle,n}\\
 			& \qquad \qquad \Big[ \left(  \mathcal{J}_{-n}(u_{ij}) \, e^{-in\omega t'} \, \hat{v}_{ij} \right)  + \left( i \leftrightarrow j \right) \Big] \lvert\Psi_{0}  \rangle_{t'}.
	 \end{aligned}
 \end{equation}
In the above equation we have used $e^{i A \sin\omega t} = \sum_{n=-\infty}^\infty \mathcal{J}_n(A) \ e^{in\omega t}$. Noting that $\vert \Psi_0 \rangle_t$ is a slow function of time, one can integrate by parts and drop the resulting integral, as being higher order in $t_h$:
 \begin{equation}
	\begin{aligned}
		\,  \lvert \Psi_{1} \rangle_t   &= - t_h \Bigg\{ \sum_{\langle ij\rangle , n} \bigg[ \left( \mathcal{J}_{-n}(u_{ij}) \, \frac{e^{-in\omega t}}{U - n\omega +  \bar{T} } \,    \hat{v}_{ij} \lvert\Psi_{0} \rangle_t \right) \\
		&\qquad \qquad \qquad+  \left( i \leftrightarrow j \right) \bigg] \quad + \mathcal{O}\left( t_h^2/U^2 \right) \Bigg\}.
	\end{aligned}
\end{equation}
The remainder is $\mathcal{O}\left( t_h^2/U^2 \right)$, because it comes from neglecting an integral which contains a factor of $1/(U - n\omega)$ and a time derivative of $\lvert\Psi_{0} \rangle_t$, which has leading order contribution proportional to $t_h/U^2$, as we will see. Note also that we are collectively showing all $ U - n\omega$ by $U$ in the argument of $\mathcal{O}$. One can plug this back into the first equation of \eqref{eq:truncated_eom} to get the following relation for the time evolution equation of $\vert \Psi_0 \rangle_t$:
\begin{equation}\label{eq:eom_psi0}
	\begin{aligned}
		i\partial_t\vert\Psi_0\rangle_t &= \sum_{\langle ij\rangle,\langle i'j'\rangle , m n} \bigg[ \left( f_{i'j'ij}^{mn}(t)\,\hat{G}_{i'j'ij}(U-n\omega)\,\vert\Psi_0\rangle_t \right) \\ 
		& \qquad \qquad \qquad + (i \leftrightarrow j) \bigg] + \left[ i' \leftrightarrow j' \right] ,
	\end{aligned}
\end{equation}
where 
\begin{equation}\label{eq:fijij_def}
	f_{i'j'ij}^{mn}(t)=- \ e^{i(m-n)\omega t}\mathcal{J}_{-n}(u_{ij})\mathcal{J}_m(u_{i'j'}),
\end{equation}
and the operator 
\begin{equation}\label{eq:capital_g_def}
	\hat{G}_{i'j'ij}(E)= t_h^2 \, \hat{P}_0 \ \hat{v}_{i'j'} \, \frac{1}{E+\bar{T}} \,\hat{v}_{ij} \ \hat{P}_0,
\end{equation} 
creates a DH pair at $(i,j)$, propagates it according to $(E + \bar{T})^{-1}$, and finally annihilates the pair at $(j',i')$.

Let us first consider a situation where $U - n \omega \gg t_h$ in equation \eqref{eq:eom_psi0}, under this condition one is able to neglect $\bar{T}$ in the operator $\frac{1}{U - n\omega - \bar{T}}$ which appears in $G(U - n\omega)$ and thus $G(U - n\omega)$ can be well approximated by $\left[ P_0 \ \hat{v}_{i'j'} \frac{1}{U - n\omega} \hat{v}_{ij} \right]$. Noting that the fraction $\frac{1}{U - n\omega}$ is a number, and that $P_0 \hat{v}_{ji} \, \hat{v}_{ij} P_0 = \left(  \frac12 - 2 \ \mathbf{S}_i\cdot\mathbf{S}_j \right) P_0,$ the evolution equation for $\vert \Psi_0 \rangle_t$ becomes:
	\begin{equation}\label{eq:shrodinger_psi_spin}
		i \ \partial_t |\Psi_0\rangle_t = - \sum_{\langle ij \rangle} \ J_{ij}(t) \left( \frac14 - \ \mathbf{S}_i\cdot\mathbf{S}_j  \right) \ |\Psi_0\rangle_{t},
	\end{equation}
with,
			 $$ J_{ij}(t) = \sum_{m,n=-\infty}^\infty  e^{i (m - n)\omega t} \mathcal{J}_m(u_{ji})\mathcal{J}_n(u_{ji}) \left(  \frac{4t_h^2}{U - n\omega}  \right).$$
		Since $J_{ij}(t)$ is periodic, the above equation of motion can be treated using the Floquet formalism. By virtue of a high frequency expansion, the Floquet effective Hamiltonian can be expanded in a power series in $1/\omega$, $H^{\text{eff}} = \sum_{n = 0}^{\infty} H_n/\omega^n$, where $H_n$ is $\mathcal{O}(J_{ij}^{n+1})$. When the driving frequency $\omega$ is much larger than	the exchange energy, the leading term is given by time averaging \cite{eckardt2015high,bukov-aip-15}:
		\begin{equation}
			H^{\text{eff}}=\frac{\omega}{2\pi}\int_0^{2\pi/\omega}dt \; \left(\sum_{\langle ij\rangle}J_{ij}(t)\,\mathbf{S}_i\cdot\mathbf{S}_j \right) + \mathcal{O}(J_{ij}^2/\omega)\;.
		\end{equation}
		Here we have dropped the constant term. With time averaging one arrives at the effective exchange parameter:
\begin{equation}
 \bar{J}_{ij} = \frac{\omega}{2\pi}\int_0^{2\pi/\omega}dt\;  J_{ij}(t)=\sum_{n=-\infty}^{\infty}\frac{4t_h^2\mathcal{J}_n^2(u_{ij})}{U - n\omega}\;.
\label{eq:time-average-exchange}
\end{equation}
This shows that the effective spin exchange interaction of the Floquet spin Hamiltonian associated with the bond $ij$ is renormalized due to the periodic driving, and becomes dependent on both the frequency and amplitude of the drive,
$J_{\langle ij \rangle}=\sum_{n=-\infty}^{\infty}4t_h^2\mathcal{J}_n^2(u_{ij})/(U-n\omega)$. Moreover, the summation over $n$ shows the contribution of all the virtual DH excitation processes which absorb/emit integer numbers of photons, and each $n-$photon process is weighted by $\mathcal{J}_n^2(u_{ij})$. 
The energy of the virtually created DH pair which absorbs/emits $n$ photons is just $U-n\omega$ because the effects of DH hoppings are neglected. Note that this is a reproduction of the results reported in previous studies \cite{floquet-exchange-nc15, floquet-chiral-sl-arxiv16, sw-bukov-prl16}.

One expects the above result to be valid up to large times \cite{abanin2017rigorous}, i.e.~(quasi-)exponential in the ratio of frequency to effective exchange energy scale, $t_h^2/U$, and not after that due to Floquet thermalization of the system; the system becomes featureless and locally indistinguishable from an infinite temperature system, due to the absorption of energy in the form of reorderings in the spin configuration of the system. 

However, the system would also be heated up by absorbing photons to create doublon-holon pairs. When the rate of  DH generations is non-negligble, the local properties of the system can no longer be captured by the low-energy spin dynamics.

The result in equation \eqref{eq:time-average-exchange} shows no imaginary part for the effective exchange interaction, hence the effective Floquet Hamiltonian dynamics in the $0$-doublon subspace shows no departure from this subspace; thus the physics of doublon-holon creation is not captured by this result. The reason for this is that the finite band-width of the virtual doublon-holon pairs in the time dependent perturbation theory is neglected by dropping $\bar{T}$ in the definition of $G$ (equation \eqref{eq:capital_g_def}) . We will study the effects that arise from restoring the finite bandwidth of these excitations next.

With the above considerations in mind, one can think of expanding $G$ in equation \eqref{eq:capital_g_def}, in a power series as follows:
\begin{widetext}
\begin{equation}\label{eq:G_series_hopping}
	G_{i'j'ij}(E) = P_0 \ v_{i'j'}\left( \frac{1}{E} \left[ 1 + \left(- \frac{\bar{T}}{E}\right) + \left(- \frac{\bar{T}}{E}\right)^2 + \left(-\frac{\bar{T}}{E}\right)^3 + \ldots \right]\right) v_{ij} \ P_0 .
\end{equation}	
\end{widetext}
The above series indicates that one should take into account all the possible virtual hopping processes taking place for the virtual doublon holon pair. 
	 The effect of considering all virtual hopping processes is two-fold.  First, it introduces subleading corrections to the evolution equation for $\vert \Psi_0 \rangle_t$ (analogue of \eqref{eq:shrodinger_psi_spin}) -- these corrections can be of the same Heisenberg-interaction form as in \eqref{eq:shrodinger_psi_spin}, and may include new forms like four-spin interactions. Note that we have already dropped comparable corrections by working up to the second order in our time dependent perturbation theory. Second and most important for our purposes, they account for the effects of a finite band width of the virtual excitations involved. To treat this in what follows we make a key assumption: we only consider the hopping processes which bring the DH pairs back to where they were created, with a final spin configuration which is identical to the initial configuration.  Taking the other \textit{nonlocal} processes into account will bring in higher order effects in the interaction terms and also the bandwidth of excitations. With the above assumption, it follows that
\begin{equation}
i\partial_t\vert\Psi_0\rangle_t=\sum_{\langle ij\rangle}\sum_{m,n=-\infty}^{\infty} \left( H_{ij}^{mn}(t)\vert\Psi_0\rangle_t \right) + \left( i \leftrightarrow j \right),
\label{eq:schordinger_time_dependent_psi0}
\end{equation}
where 
\begin{equation}\label{eq:hspin-t}
\begin{aligned}
	H_{ij}^{mn}(t)= \sum_{\sigma \sigma'} t_h^2\,f_{ij}^{mn}(t) \, c^\dagger_{j\sigma'} c_{i\sigma'}  \, c^\dagger_{i\sigma} c_{j\sigma} \ g_{\text{dh}}(U-n\omega)\;,
\end{aligned}
\end{equation}
and
$g_{\text{dh}}(E) = \!_t\langle \Psi_0\vert c^\dagger_{j\sigma} c_{i\sigma}  \, (E + \bar{T})^{-1} \, c^\dagger_{i\sigma} c_{j\sigma} \vert \Psi_0\rangle_t$ is the DH Green's function, and 
$f_{ij}^{mn}(t)$ is defined in equation \eqref{eq:fijij_def}. Note that the dependence of $g_{\text{dh}}$ on $(i,j,\sigma,\sigma' )$ and also the state $\vert \Psi_0\rangle_t$ is understood despite the notation not showing it.

\subsection{The Green's function}
To calculate the Green's function of the virtual doublon holon pair, we decompose the time-averaged DH pair hopping operator
$\bar{T}$ into a sum of a doublon-hopping and a holon-hopping term, $\bar{T}=\bar{T}_{\text{d}} + \bar{T}_{\text{h}}$, with $\bar{T}_{\text{d}}$ and $\bar{T}_{\text{h}}$ the hopping operator for the doublon and the holon in the 1-doublon subspace.
 In principle, the motions induced by $\bar{T}$ (which is the time-averaged hopping restricted to the 1-doublon subspace), for the doublons and for the holons are correlated, but following \cite{sensarma2009modulation}, we neglect the correlations between the motions of the doublons and the holons. We also add an infinitesimal negative imaginary part to the denominator of the Green's function which can be interpreted as the reciprocal of the time over which the drive is turned on.
\begin{equation}\label{eq:gdh_decomposed}
	\begin{aligned}
		g_{\text{dh}} \left( E - i\delta \right) & = \langle \Psi_0 \vert   c^\dagger_{j\sigma} c_{i\sigma}  \left( \frac{1}{E + \bar{T} - i \delta } \right) c^\dagger_{i\sigma} c_{j\sigma}\vert \Psi_0 \rangle \\
		&= - i \int \frac{d\Omega}{2\pi} \; \langle \Psi_0 \lvert c^\dagger_{j\sigma} c_{i\sigma}  \ \frac{1}{\Omega + \bar{T}_{\text{h}} - i\delta} \\
		& \qquad  \qquad \times \quad \frac{1}{E - \Omega + \bar{T}_{\text{d}} - i\delta}  \ c^\dagger_{i\sigma} c_{j\sigma}  \rvert \Psi_0 \rangle\\
		&= - i \int \frac{d\Omega}{2\pi} \; g_{\text{h}}(\Omega - i\delta)\ g_{\text{d}}(E - \Omega - i \delta),
	\end{aligned}		
\end{equation}
where $g_{\text{d}}$ and $g_{\text{h}}$ are the doublon and the holon Green's functions:
\begin{equation}\label{eq:g_dh_def}
	\begin{aligned}
		g_{\text{h}}(E-i\delta) &=  \langle \Psi_0 \lvert   c^\dagger_{j\sigma} \ \frac{1}{E + \bar{T}_{\text{h}} - i \delta}  \ c_{j\sigma} \rvert \Psi_0 \rangle,\\
		 g_{\text{d}}(E - i \delta) &= \langle \Psi_0 \lvert   c_{i\sigma} \ \frac{1}{E + \bar{T}_{\text{d}} - i \delta}  \ c^\dagger_{i\sigma} \rvert \Psi_0 \rangle.
	\end{aligned}
\end{equation}
Let us emphasize once more that in order to obtain the result of equation \eqref{eq:gdh_decomposed}, we have assumed a mean field approximation to be valid and that the motion of the doublon and that of the holon are not correlated. \cite{sensarma2009modulation}

To calculate the Green's functions defined above, we work in the retraceable path (RP) approximation proposed by Brinkman and Rice\cite{brinkman-rice-prb70}. To compute the Green's function for a holon, for example, with the same initial and final locations and spin configurations (as in \eqref{eq:g_dh_def}), one needs to consider all the hopping processes which bring the particle back to its original location, and in the meanwhile bring the spin configuration back to the original one; in the RP approximation scheme, this can be done if one takes every path that starts at the given location and terminates at the same point, with the constraint that the hopping holon should exactly retrace its forward going path in its way back to the original location. With this constraint, every spin reordering that is done in the forward going path is corrected when the particle is getting back to its original position. Note that this prescription does not capture all the possible processes; what is missing is the contribution by the paths that are closed loops and keep the final and initial spin configurations the same. As Brinkman and Rice showed, such closed loops will contribute first at order $t_h^{12}$ in the antiferromagnetic spin background for example, and are thus negligible. Note that the single doublon Green's function has the same analytical form as the holon Green's function. 

Finally $g_{\text{dh}}$ for a generic state on the right hand side of the first line of \eqref{eq:schordinger_time_dependent_psi0} can be approximated using the convolution integral in \eqref{eq:gdh_decomposed} in terms of $g_{\text{h}}$ and $g_{\text{d}}$, which are calculated using RP approximation. With this prescription, as we will see, the state dependence and also site dependence of the Green's functions $g_{\text{h}}, g_{\text{d}}$ and therefore $g_{\text{dh}}$, when non-zero, are dropped.  For a brief review of the RP approximation, we refer the reader to Appendix \ref{app:RP}.

As is shown in the appendix, after Ref.~\onlinecite{brinkman-rice-prb70}, one has the following form for the holon Green's function in the RP approximation:
\begin{equation}\label{eq:single_particle_green_RP}
	g_{\text{h}} (E) =\frac{2(z-1)}{E \left[(z-2)+z\sqrt{1-4(z-1)\bar{t}_h^2/E^2} \right] }\;,
\end{equation}
where $z$ is the coordination number $\bar{t}_h\!=\!t_h\mathcal{J}_0(u_{ij})$ is the time-averaged hopping amplitude between sites $i$ and $j$. Here, we take the polarization of the radiation such that the electric field amplitude is isotropic and thus $u_{ij}$ does not depend on the bond directions, for the sake of simplicity.
However, in principle $u_{ij}$ is different for different bonds which makes the hopping $\bar{t}_h$ anisotropic, and one needs to 
solve coupled self-energy equations for different bond directions self consistently in order to calculate the holon's Green's function. A first approximation in that case will be to average the hopping amplitudes along the different bond directions. (Please refer to Appendix \ref{app:RP} for more discussion.) This kind of treatment is also adopted in the case of multi-orbital Mott insulators as will be discussed in Sec.~\ref{sec:floquet_spin_orbital}. 

\begin{figure}[!t]
	\centering \includegraphics[width=0.45\textwidth]{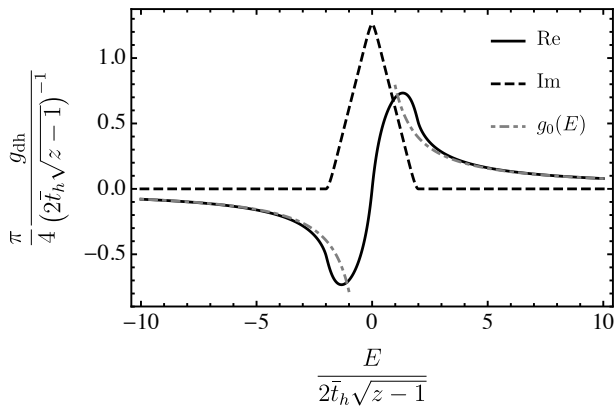}
	\caption{\label{fig:gdh_calc} Real and imaginary part of the Green's function computed with the Retraceable Path approximation. The Green's function $g_0(E) = \frac{1}{E}$ is plotted for reference. The value of $z=6$ is chosen for this plot.}
\end{figure}

It is obvious from the form in equation \eqref{eq:single_particle_green_RP} that $g_{\text{h}}(E)$ behaves as $\frac{1}{E}$ for large values of $E$. Furthermore, $g_{\text{h}}(E)$ when viewed in the complex $\tilde{E} = E/(2\bar{t}_h \sqrt{z-1})$ plane has a branch cut with the two end points $\tilde{E}=1$ and $\tilde{E} = -1 $. Note that $g_{\text{d}}$ has the same analytic expression as $g_{\text{h}}$. With the above form of the Green's function for the doublon and the holon, $g_{\text{dh}}$ is calculated as discussed in Appendix \ref{app:convolution}. A plot of $g_{\text{dh}}$ for $z=6$ obtained this way can be found in Figure \ref{fig:gdh_calc}. 

With $g_{\text{dh}}$ at hand, it is easy to derive the effective exchange parameter, similar to what was done previously:
\begin{equation}
 \bar{J}_{ij} = \sum_{n=-\infty}^{\infty} 4t_h^2\mathcal{J}_n^2(u_{ij})\ g_{\text{dh}}(U - n\omega)\;.
\label{eq:time_average_exchange_gdh}
\end{equation}
It can be seen from Figure \ref{fig:gdh_calc} that $g_{\text{dh}}(U - n \omega)$ has both real and imaginary parts when $\vert U-n\omega\vert <4\bar{t}_h\sqrt{z-1}$;  the real part contributes to the ordinary exchange parameter, while the imaginary part can be related to the dounlon-holon creation rate at bond $\langle ij \rangle$ and thus also to the increase in the local density of DH pairs. From Fig.~\ref{fig:gdh_calc} it is inferred that the imaginary part of $g_{\text{dh}}$ is always positive, this along with the fact that the effective Hamiltonian has a term like $\bar{J}_{ij}\left( -\frac14 + \ \mathbf{S}_i\cdot\mathbf{S}_j  \right)$ for every bond guarantees that the creation rate is always positive;
in fact the creation rate is zero for neighboring triplets and positive for neighboring singlets. Note that the exchange interaction parameter derived in the previous section \eqref{eq:time-average-exchange} by neglecting DH hoppings, can be consistently recovered by noting that $g_{\text{dh}}(U - n \omega)$ behaves similar to $\frac{1}{U - n \omega}$ for sufficiently large values of $\left( U - n \omega \right) / \left(\sqrt{z - 1}\, \bar{t}_h \right)$.

The above considerations show that the Floquet spin model breaks down  when the photon energy $\omega$ (setting $\hbar\!=\!1$) is in resonance with the interaction energy $U$, i.e. $n\omega$ is around $U$. In such resonant regime, the periodic driving would generate real DH pairs, and the description of the system by the low-energy spin dynamics is no longer valid. As shown in Appendix \ref{app:convolution} and discussed above, the excitation spectrum has a finite bandwidth $4\sqrt{z-1}\bar{t}_h$ due to the hoppings of the DH pairs. As a result of this, within the approximation scheme used here, real DH pairs are generated as long as the frequency is within this excitation band. On the contrary, when $\omega$ is outside the DH band, the DH creation rate is tiny and the description of the system by an effective Floquet spin Hamiltonian is still valid, but the expression of of $J_{\langle ij \rangle}$ would be modified due to the DH hoppings.

One can further find an expression for the rate of the local DH pair creation (when close to a resonance) using the above formalism. The imaginary part of $g_{\text{dh}}$, when non-zero, has a typical value of order $\frac{1}{\bar{t}_h}$, this corresponds to an imaginary effective Hamiltonian of order $t_h$, and thus a time scale for DH pair creation rate $\sim \frac{1}{\bar{t}_h}$; while on the other hand, the time scale for spin dynamics due to the effective Hamiltonian is of order $\frac{U}{\bar{t}_h^2}$. Clearly the latter is much larger in the insulating limit. Thus in order to find the DH pair creation rate, we would restrict our attention to the ground
state spin configuration of the static Hamiltonian, which is antiferromagnetic order in our case. In other words, the Floquet spin dynamics which is induced by turning on the laser radiation can safely be neglected. As pointed above, the rate of increase in the density of DH pairs $\rho_{\text{dh}}$ is basically the doublon creation rate in this spin ground state; this is nothing but the decay rate of the spin ground state calculated using the imaginary part of the Green's function described above, and thus takes the following form:
\begin{equation}\label{eq:DH_creation_rate_single}
	\begin{aligned}
	\frac{\partial}{\partial t} \rho_{\text{dh}} &= \frac{1}{N} \sum_{\langle ij \rangle} \sum_{n=-\infty}^{\infty} 4t_h^2\mathcal{J}_n^2(u_{ij})\ \mathrm{Im}\, g_{\text{dh}}(U - n\omega)\\
	& \qquad \times \qquad  \langle \Psi_0 \vert  \left[ \frac14 - \ \mathbf{S}_i\cdot\mathbf{S}_j  \right] \vert \Psi_0 \rangle
	.
	\end{aligned}
\end{equation}
where one can use the spin ground state of the static Hamiltonian for $\vert \Psi_0 \rangle$ in this relation. A consistent result can also be derived using Fermi's golden rule.

\subsection{Small frequency regime}
With the form \eqref{eq:DH_creation_rate_single} for the creation rate of DH pairs at hand, we would like to turn attention to the study of doublon creation rate at small frequencies. At very small frequencies, one expects the absorption of a high number of photons for supplying the energy needed for the creation of a DH pair, as a result of this one expects large values of $n$ (of order $U/\omega$) to only contribute to the sum in equation \eqref{eq:DH_creation_rate_single}. One can further justify this point by Noting that the function $\mathrm{Im} g_{\text{dh}}(U -n \omega)$ in \eqref{eq:DH_creation_rate_single} is nonzero when its argument is in a window around $0$ (figure \ref{fig:gdh_calc}), and thus only terms with $n \sim U/\omega$ contribute to the sum.  

Turning to the Bessel function in the sum, we note that a Bessel function of large order is essentially zero until its argument gets comparable to its order, this can be seen by checking the integral representation of a Bessel function $\mathcal{J}_n(A) = \frac{\omega}{2\pi} \int_0^{2\pi/\omega} dt \; e^{i A \sin \omega t} e^{-in\omega t}$; when $n$ is large and $A$ is not, the factor $e^{-in\omega t}$ oscillates rapidly and in one of its periods, $e^{i A \sin \omega t}$ is almost constant, this makes the integral negligible. The integral becomes not very small, only when $A$ becomes comparable to $n$. As argued above, $n$ should be of order $U/\omega$ and thus noting that the argument of the Bessel functions in \eqref{eq:DH_creation_rate_single} is $u_{ij} = eEa / \omega$, with $a$ being the lattice constant, one needs $eEa \sim U$  for a non-negligible absorption.

To present a more accurate treatment, we will focus on a given bond and assume that the spins on the two ends of the bond are aligned antiferromagnetically. The change in the local density of DH pairs due to creation of a pair at the two ends of a given bond $\langle ij\rangle$ reads%
\footnote{Since here we are dealing with frequencies that are very smaller than the energy scales of the model, it would be legitimate to use $\tilde{T}(t=0)$ in equations defining the Green's functions, such as \eqref{eq:capital_g_def} and \eqref{eq:G_series_hopping}. This means that the hopping operators used in defining Green's functions are essentially the static ones.
}:
\begin{equation}\label{eq:absorb}
	\frac{\partial}{\partial t} \rho_{\text{dh}} = \sum_{n=-\infty}^{\infty} 4t_h^2\mathcal{J}_n^2(u_{ij})\ \mathrm{Im}\, g_{\text{dh}}(U - n\omega) \times \frac{1}{2}
\end{equation}
The site indices will be suppressed in the what follows. First, we present a numerical evaluation of the sum in \eqref{eq:absorb} for small values of $\omega$, and different values of the electric field energy $\Phi = eEa = u\omega$. 

\begin{figure}[!h]
	\centering \includegraphics[width=0.47\textwidth]{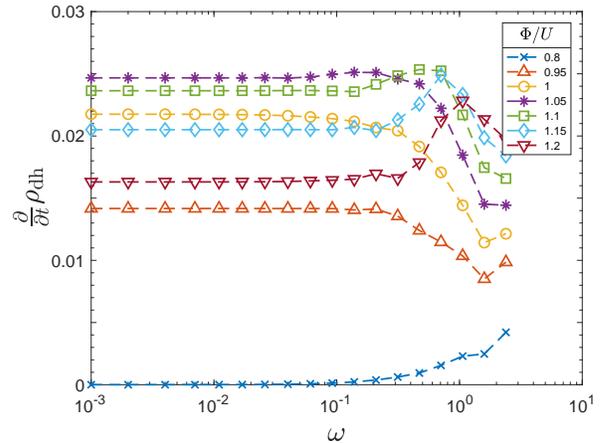}
	\caption{\label{fig:different_phi_behaviour}\footnotesize Semi-logarithmic plot of DH pair creation rate at a given bond. This plot shows how the DH creation rate behaves for small frequencies, for different values of electric field energy near $U = 10$. All energies are expressed in units of $2t_h \sqrt{z-1}$ and thus $\frac{1}{2t_h \sqrt{z-1}} \frac{\partial}{\partial t} \rho_{\text{dh}}$ versus $\frac{\omega}{2t_h \sqrt{z-1} }$ is actually plotted here.}
\end{figure}

The rate of change in the density of DH pairs given by equation \eqref{eq:absorb} is evaluated numerically for a range of small frequencies, while the electric field energy is varied. In this section, all energies are expresssed in units of $2 t_h \sqrt{z-1}$, and all times in units of its inverse. A plot of DH creation rate for $U = 10$ can be found in figure \ref{fig:different_phi_behaviour}.
Figure \ref{fig:different_phi_behaviour} shows that among many field dependent behaviors, something similar happens for different field energies, at very small frequencies; the creation rate shows a saturation for different values of field energy, i.e., a constant value is maintained over two orders of magnitude of change in frequency. This suggests the possibility for existence of a zero-frequency limit in the DH pair creation rate. 

%which can also be interpreted as the rate of breakdown of the Mott insulator.

\begin{figure}[!h]
	\centering \includegraphics[width=0.47\textwidth]{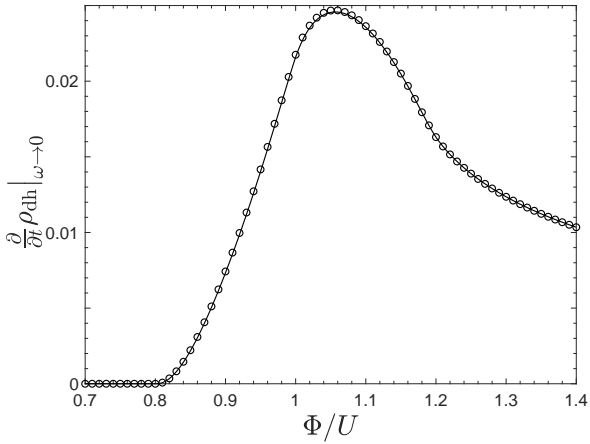}
	\caption{\label{fig:sat_values}\footnotesize  The zero-frequency limit of DH creation rate plotted as a function of electric field strength. The data points correspond to the values of zero-frequency limit, obtained by finding the saturation values at small frequencies for different field strengths numerically. The solid line on the other hand shows the prediction of the analytical result \eqref{eq:final_form_S}. This shows that there is a very good agreement between the two results.}
\end{figure}

The small frequency saturation value can be extracted numerically for different electric field strengths, a plot of which is presented in figure \ref{fig:sat_values} as the data points. First, this plot shows that a nonvanishing zero-frequency limit exists only if $\Phi/U \geq 0.8$, i.e., when the electric field energy is above  the lower bound of the DH excitations, and below this value it is negligible (zero within our approach). Second, it also shows that the maximum zero-frequency limit of the DH creation rate occurs at an electric field strength just slightly higher than $U$. This zero-frequency limit of DH creation rate can also be interpreted as the rate for the breakdown of a Mott insulator, when exposed to a static electric field and thus we indeed expect from figure \ref{fig:sat_values} that the maximum static breakdown rate happens when $\Phi$ is very close to $U$.

We then turn our focus to an analytical study of the asymptotic behavior of the sum in equation \eqref{eq:absorb}. Since the imaginary part of the Green's function $g_{\text{dh}}(E)$ is only nonzero when its argument sits in the window $(U - 2 \, , \, U + 2)$, the sum can be rewritten as:
\begin{equation}\label{eq:sum_bessel}
	\begin{aligned}
		\mathcal{S} &= \sum_{n=-\infty}^\infty \mathcal{J}_n^2(\Phi/\omega)\, \mathrm{Im} g_{\text{dh}}(U - n \omega) \\
	&= \sum_{n \sim U/\omega - 2 /\omega}^{U/\omega + 2 /\omega}  \mathcal{J}_n^2(\Phi/\omega) \, \mathrm{Im} g_{\text{dh}}(U - n \omega).
	\end{aligned}
\end{equation}
In the limit of very small frequency, $n$ is a large number for all the terms in the above sum, and thus one can use an asymptotic form for Bessel functions of high order (Appendix \ref{app:asymptotic}). After substituting the Bessel functions with the asymptotic forms, the sum can further be converted to an integral, and thus finally the quantity $\mathcal{S}$ in the zero-frequency limit takes the form (Appendix \ref{app:asymptotic}):
\begin{equation}\label{eq:final_form_S}
	\lim_{\omega \to 0}\mathcal{S} = \frac{1}{\pi} \int_{U - 2}^{\Phi} d\bar{\nu} \; \left(\frac{1}{\left(\frac{\Phi}{\bar{\nu} }\right)^2 - 1}\right)^{1/2}  \frac{1}{\bar{\nu} } \  \mathrm{Im} g_{\text{dh}}(U - \bar{\nu}).
\end{equation}

This analytical form for the zero-frequency limit is plotted and compared to the numerical result in figure \ref{fig:sat_values} as the solid line, and it can be seen that there is a very good agreement between this analytical result and the saturation values found numerically. Indeed, this shows that there is a zero-frequency limit for the creation rate of DH pairs and thus as mentioned above, the breakdown rate of the Mott insulator due to a static electric field, can be read from figure \ref{fig:sat_values}. These results can be easily generalized to the multi-orbital Mott insulators by the treatments introduced in section \ref{subsec:DH_multi}. 

\section{Floquet Spin-Orbital Model}\label{sec:floquet_spin_orbital}

\subsection{Time Dependent Perturbation Theory}

The previous discussion of the periodically driven Hubbard Model can be generalized to the case of periodically driven
multi-orbital Mott insulators with Kanamori local interactions \cite{kanamori-63}
\begin{align}
H_{\textrm{K}}=&U \, \sum_{i,\alpha}\hat{n}_{i\alpha\uparrow}\hat{n}_{i\alpha\downarrow}+
U'\sum_{i,\alpha <\beta,\sigma,\sigma'}\hat{n}_{i\alpha\sigma}\hat{n}_{i\beta\sigma'}\;\nn
&-J_{\textrm{H}}\sum_{i,\alpha < \beta,\sigma,\sigma'}c^{\dagger}_{i\alpha\sigma}
c_{i\alpha\sigma'}c^{\dagger}_{i\beta\sigma'}c_{i\beta\sigma}\;\nn
&+J_{\textrm{P}}\sum_{i,\alpha < \beta,\sigma}
c^{\dagger}_{i\alpha\sigma}c^{\dagger}_{i\alpha-\sigma}c_{i\beta\sigma}c_{i\beta -\sigma}\;,
\label{eq:kanamori}
\end{align}
where $U$ and $U'$ are the intra-orbital and inter-orbital direct Coulomb interactions. $J_{\textrm{H}}$ and $J_{\textrm{P}}$ denote the onsite exchange interaction (Hunds' coupling) and the pair hopping respectively; the sets of indices $\{i,j\}$, $\{\alpha,\beta\}$, $\{\sigma,\sigma'\}$ denote the lattice sites, orbitals and spin degrees of freedom. As in the case of Hubbard model, the effect of the laser radiation is manifested in the kinetic energy via the so called Peierls substitution,
\begin{equation}
T_t=\sum_{\langle ij \rangle, \alpha\beta,\sigma} \left( t_{ i \alpha, j\beta} \ e^{iu_{ij}\sin\omega t} \ c^{\dagger}_{i\alpha\sigma}c^{\vphantom\dagger}_{j\beta\sigma} + \textrm{h.c.} \right) ,
\label{eq:multi-hopping}
\end{equation}
where $t_{i\alpha,j\beta}$ represents the hopping amplitude from orbital $\beta$ at site $j$ to orbital $\alpha$ at site $i$. Note that Hermiticity dictates $t_{i\alpha,j\beta} = t_{j\beta,i\alpha}^*$, along with $u_{ij} = - u_{ji}$.

In the multi-orbital case, we also need to consider the crystal-field splittings ($H_{\textrm{CF}}$). 
In addition to the giant $t_{2g}-e_g$ splitting  in typical perovskite transition-metal oxides, 
there may be additional splittings within
the $t_{2g}$ and/or $e_g$ manifold due to the octahedral rotations, tiltings \cite{okanderson-prl04} and 
Jahn-Teller distortions \cite{kugel1982jahn}. The crystal field splitting between $t_{2g}$ and $e_g$ levels
is on the order of a few eV, but the splittings due to octahedral rotations and Jahn-Teller distortions are typically
much smaller. Throughout this paper we only consider the $t_{2g}$ orbitals. Within the quasi-degenerate 
$t_{2g}$ levels we further include the crystal-field splittings  from various octahedral distortions and tiltings,
\begin{equation}
H_{\textrm{CF}}=\sum_i\sum_{\alpha,\beta,\sigma}\epsilon_{i,\alpha\beta}\,c^{\dagger}_{i\alpha\sigma}c^{\vphantom\dagger}_{i\beta\sigma}\;.
\end{equation}
Including all these terms, we find the total periodically driven Hamiltonian as $H_t=T_t + H_K + H_{\textrm{CF}}$. 

We consider the limit that the typical interaction energy
scale (a few $\mathrm{eV}$) is much greater than the hopping energy scale ($\sim\!0.1\,\mathrm{eV}$ in $3d$ transition-metal oxides), 
and consider $T_t$ as a perturbation to $H_{\textrm{K}}$. In the non-driven case, the low-energy physics is dominated by the spin and orbital dynamics, which is well described by the Kugel-Khomskii \cite{kugel1973crystal,kugel1982jahn} and similar spin-orbital models, and can be derived
using time independent second-order perturbation theory.
Here we derive a time-dependent spin-orbital model for periodically driven multi-orbital Mott insulators using time-dependent perturbation theory. We consider the situation of one electron occupancy at every site in the ground state of the static system. The key difference with the single-orbital case is that here, the virtual doubly-occupied states have different energies for different spin and orbital configurations. We also neglect the $H_{\text{CF}}$ in the calculation of energies of the virtual double-occupancy states, since $H_{\text{CF}}$ has a much smaller energy scale than $H_{\text{K}}$.

To find an effective equation of motion in this model, in an approach similar to the one used in the single orbital case, we
expand a generic state of the system in terms of the states with different numbers of double occupancies. We then truncate the equations of motion similarly to arrive at:
 \begin{equation}\label{eq:truncated_eom_multi}
	\begin{aligned}
		i \partial_t |\Psi_0\rangle_t &=  \hat{P}_0 \ T_t |\Psi_1\rangle_t ,\\
		i \partial_t |\Psi_{1}\rangle_t &=  H_{\text{K}} \ |\Psi_{1}\rangle_t + \ T_t |\Psi_0\rangle_t +  \tilde{T}_t |\Psi_1\rangle_t .
	\end{aligned}
\end{equation}
where $\tilde{T} = \hat{P}_1 T_t \hat{P}_1$, and the hopping operator $T_t$ in this case is multi-orbital.
First, we consider the case where the effect of hopping of virtual excitations is negligible, which occurs when one is sufficiently away from a resonance. In this case, one can safely drop the term $\hat{P}_1 \ T_t |\Psi_1\rangle_t$ in the second equation of \eqref{eq:truncated_eom_multi}, and with the same manipulations done in the single orbital case, one arrives at the following form for the effective equation of motion of $|\Psi_0\rangle_t$:
\begin{widetext}
\begin{equation}
	\begin{aligned}
		i \partial_t |\Psi_0\rangle_t 
		& = - \sum_{n,m} \sum_{\langle ij \rangle}  \mathcal{J}_n \left(u_{ji}\right) \mathcal{J}_m \left(u_{ji}\right) \,  e^{i (m - n) \omega t}  \sum_{\alpha\beta\sigma\alpha'\beta'\sigma'} t_{i\alpha j\beta} t_{j\alpha' i\beta'} \ c^\dagger_{j\alpha'\sigma'} c^{\vphantom\dagger}_{i\beta'\sigma'} \, \frac{1}{H_{\text{K}} - n \omega} \, c^\dagger_{i\alpha\sigma} c^{\vphantom\dagger}_{j\beta\sigma} \ |\Psi_0\rangle_t  + \ \left( i \leftrightarrow j \right),
	\end{aligned}
\end{equation}	
\end{widetext}
where $H_{\textrm{K}}$ is the Kanamori interaction operator defined in \eqref{eq:kanamori}.

The Floquet spin-orbital effective Hamiltonian can now be obtained by time-averaging:
\begin{equation}
	H^{\text{eff}} = \sum_{n} \sum_{\langle ij \rangle} \mathcal{J}^2_n \left( u_{ij} \right) \hat{\Lambda}_{ij} (n \omega),
\end{equation}
with $\hat{\Lambda}_{ij}$ defined as:
\begin{equation}
	\begin{aligned}
		\hat{\Lambda}_{ij}  (n \omega)	&= - \Bigg[ \sum \Big(t_{i\alpha j\beta} t_{j\alpha' i\beta'}  \ c^\dagger_{j\alpha'\sigma'} c^{\vphantom\dagger}_{i\beta'\sigma'} \, \frac{1}{H_{\text{K}} - n \omega} \\
		& \qquad \qquad \qquad c^\dagger_{i\alpha\sigma} c_{j\beta\sigma}  \Big) + \left( i \leftrightarrow j \right) \Bigg] \hat{P}_0,
	\end{aligned}
\end{equation}
with the summation done over the set of indices $\{\alpha,\beta,\sigma,\alpha',\beta',\sigma'\}$. To calculate the operator $\hat{\Lambda}_{ij}$, one should note that it creates a doublon-holon pair at sites $i,j$, then acts on the resulting state with the inverse Kanamori Hamiltonian, and annihilates the pair finally. It is a 0-doublon to 0-doublon operator and thus can be written in terms of spin and orbital operators acting on the 0-doublon subspace. As mentioned above we will compute this operator for the case of three orbitals.

Noting that $H_{\text{K}}$ has four distinct eigenvalues $E_\text{K,1} = U' - J_{\text{H}} \, , \, E_\text{K,2} = U' + J_{\text{H}} \, , \, E_\text{K,3} = U- J_{\text{P}} \, , \, E_\text{K,4} = U + 2 J_{\text{P}}$, one is able to expand $\hat{\Lambda}_{ij}  (n \omega)$ as follows
\begin{equation}
	\begin{aligned}
		\hat{\Lambda}_{ij} (n \omega) &=   \frac{\hat{\Lambda}_{ij,1}}{U' - J_{\text{H}} - n \omega}  + \frac{\hat{\Lambda}_{ij,2}}{U' + J_{\text{H}} - n \omega}\\
		& + \frac{\hat{\Lambda}_{ij,3}}{U- J_{\text{P}} - n \omega} + \frac{\hat{\Lambda}_{ij,4}}{U + 2 J_{\text{P}} - n \omega} ,
	\end{aligned}
\end{equation}
where $\hat{\Lambda}_{ij,m}$ is the spin-orbital operator corresponding to eigenvalue number $m$. One can further decompose these operators into spin and orbital parts as follows:
\begin{equation}\label{eq:Lambda_def}
	\begin{aligned}
		\hat{\Lambda}_{ij,1} &= \left(\frac{3}{4} + \mathbf{S}_i\cdot\mathbf{S}_j \right) \left[\hat{\gamma}_{ij,2} - \hat{\gamma}_{ij,1}\right],\\
		\hat{\Lambda}_{ij,2} &= \left(- \frac{1}{4} + \mathbf{S}_i\cdot\mathbf{S}_j \right)  \left[\hat{\gamma}_{ij,2} + \hat{\gamma}_{ij,1} - \hat{\gamma}_{ij,3}\right],\\
		\hat{\Lambda}_{ij,3} &= \left(- \frac{1}{4} + \mathbf{S}_i\cdot\mathbf{S}_j \right) \left[\hat{\gamma}_{ij,3} - \hat{\gamma}_{ij,4} \right],\\
		\hat{\Lambda}_{ij,4} &= \left(- \frac{1}{4} + \mathbf{S}_i\cdot\mathbf{S}_j \right) \hat{\gamma}_{ij,4} .
	\end{aligned}
\end{equation}
In the above equations, $\gamma_{ij,m}$ are the following orbital operators:
\begin{equation}\label{eq:gamma_orbital_definition}
	\begin{aligned}
		\hat{\gamma}_{ij,1} &= \sum_{\alpha_i\beta_i\beta_j} 	\hat{A}^i_{\alpha_i \beta_i} \ t_{i\alpha_i j \beta_j} \, t_{j \beta_j i \beta_i} + (i \leftrightarrow j), \\
		\hat{\gamma}_{ij,2} &= 2 \sum_{\alpha_i\beta_i\alpha_j\beta_j} 	\hat{A}^i_{\alpha_i \beta_i} \hat{A}^j_{\alpha_j \beta_j} \ t_{i\alpha_i j \beta_j} \, t_{j \alpha_j i \beta_i}, \\
		\hat{\gamma}_{ij,3} &= 2 \sum_{\alpha_i\beta_i\alpha_j} 	\hat{A}^i_{\alpha_i \beta_i} \hat{A}^j_{\alpha_j \alpha_j} \ t_{i\alpha_i j \alpha_j} \, t_{j \alpha_j i \beta_i} + (i \leftrightarrow j), \\
		\hat{\gamma}_{ij,4} &= \frac23 \sum_{\alpha_i\beta_i\alpha_j\beta_j} 	\hat{A}^i_{\alpha_i \beta_i} \hat{A}^j_{\alpha_j \beta_j} \ t_{i\alpha_i j \alpha_j} \, t_{j \beta_j i \beta_i} + (i \leftrightarrow j) .
	\end{aligned}
\end{equation}
We have introduced the orbital operators $\hat{A}^i_{\alpha_i \beta_i} = \sum_\sigma c^\dagger_{i\alpha_i \sigma} c_{i\beta_i\sigma} \, \hat{P}_0^i$, with $\alpha_i,\beta_i = 1,2,3$, as a basis for orbital operations at each site $i$, that contains only one electron. Note that $\hat{P}_0^i$ is the projector onto the states with one electron at site $i$. With the above manipulations one is able to derive an effective exchange $\hat{J}_{ij}$, which is an orbital operator in the present case:
\begin{equation}\label{eq:general_exchange_multi_noresonance}
	\begin{aligned}
		\hat{J}_{ij} = & \sum_{n=-\infty}^\infty \mathcal{J}_n^2 (u_{ij}) \\
		& \times \qquad \Bigg\{  \frac{\hat{\gamma}_{ij,2} - \hat{\gamma}_{ij,1}}{U' - J_{\text{H}} - n \omega} + \frac{\hat{\gamma}_{ij,2} + \hat{\gamma}_{ij,1} - \hat{\gamma}_{ij,3}}{U' + J_{\text{H}} - n \omega}  \\
		&\qquad \qquad + \frac{\hat{\gamma}_{ij,3} - \hat{\gamma}_{ij,4}}{U- J_{\text{P}} - n \omega}  + \frac{ \hat{\gamma}_{ij,4} }{U + 2 J_{\text{P}} - n \omega} \Bigg\}
	\end{aligned}
\end{equation}

It can be seen from \eqref{eq:Lambda_def} that the last three contributions only arise when the adjacent spins at $\langle ij \rangle$ are in a singlet state, this means that the virtual processes responsible for these terms only occur when the state of the adjacent spins is a singlet.

It is also worthwhile to study the special case $U'=U-J_{\textrm{H}}$ and $J_{\textrm{P}}\!=\!0$ \cite{georges-review13}. With such an assumption $H_{\textrm{K}}$ is rotationally invariant and there are only two distinct multiplet energy levels: $E_{\textrm{singlet}}=U$ for spin singlets, and $E_{\textrm{triplet}}=U-2J_{\textrm{H}}$ for spin triplets \cite{georges-review13}. Indeed, with this assumption $E_\text{K,1} = U - 2 J_\text{H}$ and $E_\text{K,2} = E_\text{K,3} = E_\text{K,4} = U$, and the effective exchange operator becomes:
\begin{equation}\label{eq:special_case_exchange_multi_noresonance}
	\begin{aligned}
		\hat{J}_{ij} = \sum_{n=-\infty}^\infty \mathcal{J}_n^2 (u_{ij}) \left(  \frac{\hat{\gamma}_{ij,2} - \hat{\gamma}_{ij,1}}{U - 2 J_{\text{H}} - n \omega}  + \frac{\hat{\gamma}_{ij,2} + \hat{\gamma}_{ij,1} }{U - n \omega}   \right).
	\end{aligned}
\end{equation}

\subsection{Bandwidth of Excitations}\label{subsec:DH_multi}

Now we will take into account the effects of the multi-orbital doublon-holon bandwidth.  This is quite complex in comparison to the single-orbital Hubbard model and we will consequently make a number of simplifying assumptions in order to obtain a tractable result.  While these approximations are not fully controlled, we believe they do not qualitatively affect the results.  First, we specialize in this part to the case $U'=U-J_{\textrm{H}}$ and $J_{\textrm{P}}\!=\!0$, and note again that in this case there are only two different eigenvalues for the Kanamori Hamiltonian corresponding to singlet and triplet virtual states: $E_{\textrm{singlet}}=U$, and $E_{\textrm{triplet}}=U-2J_{\textrm{H}}$.

We have to consider the multi-orbital equations of motion \eqref{eq:truncated_eom_multi} once more and this time we will not neglect the hopping term for the excitations, to see the effect of finite band-width of excitations in the multi-orbital model. To this end, we expand the 1-doublon component as $\vert\Psi_1\rangle_t = \vert\Psi_{1}^{\textrm{s}}\rangle_t + \vert\Psi_{1}^{\textrm{t}}\rangle_t$, where $\vert\Psi_{1}^{\textrm{s}}\rangle$ and $\vert\Psi_{1}^{\textrm{t}}\rangle$ denote the single-doublon states with their doublon in a spin singlet state and a spin triplet state.  As discussed before, we neglect the excited states with more than one doublon. The equations of motion can be written as:
 \begin{equation}\label{eq:truncated_eom_multi_t_s}
	\begin{aligned}
		i \partial_t |\Psi_0\rangle_t &=  \hat{P}_0 \ T_t \left( \vert\Psi_{1}^{\textrm{s}}\rangle_t + \vert\Psi_{1}^{\textrm{t}}\rangle_t \right) ,\\
		i \partial_t |\Psi_{1}^\s\rangle_t &=  U \ |\Psi_{1}^\s\rangle_t + \hat{P}_1^\s \ T_t |\Psi_0\rangle_t \\
		& \qquad \qquad + \  \tilde{T}_t^{\s\s} |\Psi_1^\s\rangle_t +  \tilde{T}_t^{\s\t} |\Psi_1^\t\rangle_t, \\
		i \partial_t |\Psi_{1}^\t\rangle_t &=  \left( U-2J_{\textrm{H}} \right) \ |\Psi_{1}^\t\rangle_t + \hat{P}_1^\t  \ T_t |\Psi_0\rangle_t \\
		& \qquad \qquad + \ \tilde{T}_t^{\t\t} |\Psi_1^\t\rangle_t + \tilde{T}_t^{\t\s} |\Psi_1^\s\rangle_t.
	\end{aligned}
\end{equation}
Here $\hat{P}_{1}^{\textrm{t}}$ and $\hat{P}_{1}^{\textrm{s}}$  are the triplet and singlet projection operators. The hopping operators are defined as $\tilde{T}_t^{ab} = \hat{P}_1^a \tilde{T}_t \hat{P}_1^b$ ($a, b=\textrm{s}, \textrm{t}$). We continue by replacing $\tilde{T}_t^{\t\t}$ and $\tilde{T}_t^{\s\s}$ by their time averages, similar to the single orbital case, and also by neglecting the two cross hoppings $\tilde{T}_t^{\s\t}$ and $\tilde{T}_t^{\t\s}$ (see Appendix \ref{app:time_averaging}). One is now able to write down the 1-doublon components in terms of the 0-doublon component at arbitrary time, and through manipulations similar to those in the single orbital case one arrives at the following form for the time evolution equation of $\vert \Psi_0 \rangle_t$ in the multi-orbital case:
\begin{equation}
\begin{aligned}
i\partial_t &\vert\Psi_0\rangle_t \\
 =& \sum_{\langle ij\rangle\langle i'j'\rangle,mn,a} \bigg[\left(
f^{mn}_{i'j'ij}(t) \ \hat{G}^{a}_{i'j'ij}(n\omega)\; \vert\Psi_0\rangle_t\right)\\
&  \qquad \qquad \qquad + (i \leftrightarrow j) \bigg] + [i' \leftrightarrow j'] \\
& \qquad + H_{\textrm{CF}} \, \vert\Psi_0\rangle_t\;,
\label{eq:h_time_multi}
\end{aligned}
\end{equation}
where $f_{i'j'ij}^{mn}(t)=-e^{i(m-n)\omega t}\mathcal{J}_m(u_{i'j'})\mathcal{J}_{-n}(u_{ij})$, and the index $a$ in the sum runs over $\{\textrm{s},\textrm{t}\}$. Note that we did not include the $H_{\text{CF}}$ term in equation \eqref{eq:truncated_eom_multi_t_s}, as we are neglecting it compared to $H_{\text{K}}$, but it has been included in \eqref{eq:h_time_multi}. The operator $\hat{G}^{a}_{i'j'ij}(n\omega)$ is defined as
\begin{equation}\label{eq:multi_green}
	\begin{aligned}
\hat{G}^{\textrm{s}}_{i'j'ij} (n\omega) = &\sum_{\alpha\beta\alpha'\beta',\sigma\sigma'}\left(t_{i\alpha,j\beta}\, t_{i'\beta',j'\alpha'}\right)\, \hat{P}_0 \,c^{\dagger}_{i'\beta'\sigma'} c_{j'\alpha'\sigma'}\times\, \\
&\; (U-n\omega + \bar{T}^{\textrm{ss}})^{-1} \, \hat{P}_{1}^{\textrm{s}}\, \,c^{\dagger}_{i\alpha\sigma} c_{j\beta\sigma}\, \hat{P}_0\, , \\
\hat{G}^{\textrm{t}}_{i'j'ij} (n\omega) =&\sum_{\alpha\beta\alpha'\beta',\sigma\sigma'} \left( t_{i\alpha,j\beta}\, t_{i'\beta',j'\alpha'}\right) \, \hat{P}_0 \,c^{\dagger}_{i'\beta'\sigma'}c_{j'\alpha'\sigma'}\times\,\\
&\; (U - 2J_{\textrm{H}}-n\omega + \bar{T}^{\textrm{tt}})^{-1}\, \hat{P}_{1}^{\textrm{t}}\,c^{\dagger}_{i\alpha\sigma}c_{j\beta\sigma}\, \hat{P}_0.
\end{aligned}
\end{equation}
Similar to the single-orbital case, we also make the following assumption for the operators $\hat{G}^{\textrm{s}}_{i'j'ij}$ and $\hat{G}^{\textrm{t}}_{i'j'ij}$: we only consider the hopping processes
which create DH pairs at the given sites, propagate them around and bring them to their initial positions and annihilate them, with a final \textit{spin-orbital} configuration which is identical to the initial configuration.  Under this assumption, equation \eqref{eq:multi_green} can be expressed as
\begin{equation}\label{eq:multi_green2}
\begin{aligned}
\hat{G}^{a}_{jiij} (n\omega) = & \sum_{\alpha\beta\alpha'\beta',\sigma\sigma'} t_{i\alpha,j\beta} \, t_{j\beta',i\alpha'} \ g_{\mathrm{dh}}^{a}(U^a - n\omega) \\
& \qquad \qquad \hat{P}_0 \, c^{\dagger}_{j\beta'\sigma'} c_{i\alpha'\sigma'} \, \hat{P}_1^a \, c^{\dagger}_{i\alpha\sigma} c_{j\beta\sigma} \, \hat{P}_0,
\end{aligned}
\end{equation}
with $a$ being either $\s$ or $\t$, with $U^\s = U $ and $U^\t = U-2J_{\textrm{H}}$ and where
\begin{equation}\label{eq:multi_green_final}
\begin{aligned}
g_{\mathrm{dh}}^{\textrm{s}} &( U - n \omega ) = \langle\Psi_0\vert c_{j\beta\sigma}^{\dagger}c_{i\alpha\sigma} \\ 
 & \qquad \times \quad \frac{\hat{P}_{1}^{\textrm{s}}}{U-n\omega+\bar{T}^{\textrm{ss}}}
c_{i\alpha\sigma}^{\dagger}c_{j\beta\sigma}\vert\Psi_0\rangle,\\
g_{\mathrm{dh}}^{\textrm{t}}  &( U - 2 J_{\textrm{H}} - n \omega )  =\langle\Psi_0\vert c_{j\beta\sigma}^{\dagger}c_{i\alpha\sigma} \\ 
 & \qquad \times \quad \frac{\hat{P}_{1}^{\textrm{t}}}{U-2J_{\textrm{H}}-n\omega+\bar{T}^{\textrm{tt}}}
c_{i\alpha\sigma}^{\dagger}c_{j\beta\sigma}\vert\Psi_0\rangle\;.
\end{aligned}
\end{equation}

In the multi-orbital case, $g_{dh}^{\textrm{t}(\textrm{s})}$ is calculated using the analogues of equations \eqref{eq:gdh_decomposed} and \eqref{eq:g_dh_def}, assuming the motions of the doublons and holons are uncorrelated:
\begin{equation}\label{eq:convulution_integral_multi}
	g_{\text{dh}}^{\t(\s)} \left( E - i\delta \right) = - i \int \frac{d\Omega}{2\pi} \; g_{\text{h}}(\Omega - i\delta)\ g_{\text{d}}^{\t(\s)}(E - \Omega - i \delta),	
\end{equation}
where $ g_{\text{h}}$ and $ g_{\text{d}}^{\t(\s)}$ are the holon and the doublon Green's functions which are defined in a similar fashion to the single-orbital case:
\begin{equation}\label{eq:g_dh_def_multi}
	\begin{aligned}
		g_{\text{h}}(E-i\delta) &=  \langle \Psi_0 \lvert   c^\dagger_{j\beta\sigma} \ \frac{1}{E + \bar{T}_{\text{h}} - i \delta}  \ c_{j\beta\sigma} \rvert \Psi_0 \rangle,\\
		 g^{\t(\s )}_{\text{d}}(E - i \delta) &= \langle \Psi_0 \lvert   c_{i\alpha\sigma} \ \frac{\hat{P}_{1d}^a}{E + \bar{T}^{\t(\s )}_{\text{d}} - i \delta}  \ c^\dagger_{i\alpha\sigma} \rvert \Psi_0 \rangle,
	\end{aligned}
\end{equation}
where $\hat{P}_{1d}^a$ projects onto subspace with zero holon and one $a$-type doublon and like the single orbital case, we have assumed a decompostion for the hopping operator $\bar{T}^{aa} = \bar{T}^a_{\text{d}} + \bar{T}_{\text{h}}$.  We will try to compute the doublon and the holon Green's functions using the retraceable path approximation in a similar approach to the one presented in section \ref{sec:floquet_spin}. One should note that in the multi-orbital case  $ g_{\text{h}}$ and $ g_{\text{d}}$ defined above, unlike the single orbital model, highly depend on the state in question. In other words, since there are orbital degrees of freedom, even with the RP approximation $ g_{\text{h}}$ and $ g_{\text{d}}^{\t(\s)}$ do not turn out to be independent of the state $\vert \Psi_0 \rangle$ since in general hopping parameters can be different for hopping events between different initial and final orbitals. 

Noting the above fact, we work in a limit that the crystal field splitting (within the $t_{2g}$ or $e_g$ manifold) is much larger than the intersite exchange energy, so that the occupied orbital at each site is uniquely determined by the crystal field term and is denoted by the orbital index $\alpha = 1$.  We would like to consider an effective holon hopping parameter between two adjacent lattice sites $i$ and $j$ as an input to our Green's functions calculated using RP approximation; in the classical-orbital regime discussed above, it is legitimate to introduce an effective hopping which accounts for hoppings from orbitals $\vert 1\rangle_{i(j)}$ to $\vert \alpha\rangle_{j(i)}$ and then back, which is denoted as $t_{ij}^{\text{eff}}$:
\begin{equation}
(t^{\text{eff}}_{ij})^2=\frac{1}{2}\sum_{\alpha}(\left| t_{i1,j\alpha}\right|^2+
\left| t_{j1,i\alpha}\right| ^2).
\label{eq:multi_eff_hopping}
\end{equation}
In a semiclassical approximation the doublon effective hopping for both of the cases ($\s , \t$) is also taken to be equal to the above value. 
The effective hopping defined above is anisotropic along different bond directions in general and furthermore, time-averaging ($\bar{T}^{\textrm{ss}(\textrm{tt})}$) introduces a factor of $\mathcal{J}_{0}(u_{ij})$ into the hopping amplitudes. For simplicity we further average over the hopping amplitudes along different bond directions, resulting in an effective isotropic DH hopping parameter for a given site $i$: 
\begin{equation}
\bar{t}_i = \frac{1}{z}\sum_{j\, \in \, \textrm{n.n.}}t^{\text{eff}}_{ij}\, \mathcal{J}_0(u_{ij}),
\end{equation} 
where ``n.n." is the abbreviation for nearest neighbor. One calculates the Green's functions on the right hand side of \eqref{eq:convulution_integral_multi}, with this value of hopping: with this prescription the DH Green's functions for both singlet and triplet cases, i.e.~$g_{\text{dh}}^{\t}$ and $g_{\text{dh}}^{\t}$, will have the same form, and this form agrees with the one discussed in the single orbital case, except for a different effective hopping here; the imaginary and real parts of such Green's function can be found in figure \ref{fig:gdh_calc}.

With all this at hand, using manipulations similar to those leading to \eqref{eq:general_exchange_multi_noresonance} and \eqref{eq:special_case_exchange_multi_noresonance}, one is able to write down the effective Hamiltonian describing the dynamics of the 0-doublon subspace for the multi-orbital case which includes the effect of bandwidth of excitations:
\begin{equation}
	\begin{aligned}
		H^{\text{eff}} &= \sum_{\langle ij \rangle, n} 
		  \bigg[ \mathcal{J}_n^2 (u_{ij})  \left( \hat{\gamma}_{ij,2} + \hat{\gamma}_{ij,1} \right) \, g_{\mathrm{dh}}^{\textrm{s}} ( U - n \omega ) \\
		 & \qquad\qquad \times \left(-\frac{1}{4} + \mathbf{S}_i\cdot\mathbf{S}_j \right)\\
	 	 & +  \mathcal{J}_n^2 (u_{ij})  \left( \hat{\gamma}_{ij,2} - \hat{\gamma}_{ij,1} \right) \, g_{\mathrm{dh}}^{\textrm{t}} ( U -2 J_{\textrm{H}} - n \omega ) \\
		 & \qquad\qquad \times \left(\frac{3}{4} + \mathbf{S}_i\cdot\mathbf{S}_j \right) \bigg],
	\end{aligned}	
\end{equation}
in which, we have used the orbital operators defined in \eqref{eq:gamma_orbital_definition}.

It is worthwhile here to make connection with our result for the multi-orbital case when the frequency is away from resonances, i.e.~equation \eqref{eq:special_case_exchange_multi_noresonance}. The functional form for the Green's functions used in the above effective Hamiltonian can be seen in figure \ref{fig:gdh_calc};  as we discussed for the single orbital case, when the argument of the Green's function is much larger than the effective hopping, and hence one is away from a resonance, the Green's functions $g_{\text{dh}}^a (U^a - n\omega)$ look very similar to $1/(U^a - n\omega)$, and thus one recovers the previous form \eqref{eq:special_case_exchange_multi_noresonance}. Furthermore, when this argument is close enough to zero, or more precisely in a window of width of the same order as the effective hopping, one expects to see nonzero imaginary part for the Green's function; this can happen when one of the excitation energies is close enough to a multiple of the frequency. Because the Green's funtion is complex, the above effective Hamiltonian becomes non-Hermitian and thus the effective evolution of  $\vert\Psi_0 \rangle_t$  becomes non-unitary; the stronger this non-unitarity becomes, the more doublons are created. Indeed, a DH pair creation rate for the multi-orbital case can also be derived given a spin-orbital configuration which will look similar to the one derived for the single orbital case \eqref{eq:DH_creation_rate_single}.

In order for the Floquet engineering of the spin-orbital dynamics to be relevant, one needs to be in a regime where not many doublons are created and thus one needs to avoid certain ranges of parameters in which the non-unitary of the effective Hamiltonian results in a large rate of doublon creation. Let us consider a concrete example to show how one can study this quatitatively: as mentioned before, the above effective Hamiltonian is applied to the orthorhombic titanates YTiO$_3$ and LaTiO$_3$ using first-principles calculations in \cite{liu2018floquet}. There, we show that the effective exchange interaction for neighboring sites can be engineered to a high degree but the effective exchange parameter turns out to be a complex number; this complex exchange parameter can be interpreted as follows, its real part shows the strength of the physical exchange interaction between neighboring sites and its imaginary part quantifies the rate of change in DH pairs density. Therefore and as we discuss in length in that letter, one will be interested in regimes where the real part of the effective exchange parameter is much larger than its imaginary part; this results in a much smaller time scale for the physical exchange interaction dynamics than the time scale for DH creation; this leads the exchange interaction to be the dominant physical effect in such a setting.

We believe that the above formalism with slight modifications can be applied to many other realistic situations involving driven multi-orbital Mott insulators. In this formalism, not only the effect of Floquet engineering can be computed, but also the rate for a channel of heating, i.e.~DH pair creation, can be quantitatively evaluated.

\section{Conclusion}
In this work, we have studied the effects of coherent laser driving of single-orbital and multi-orbital Mott insulators. 
We have developed the formalism and methods for calculating the Floquet spin and spin-orbital Hamiltonians in single-orbtial and multi-orbital Mott insulators. We have shown that the effective exchange interactions are not only dependent on the intrinsic properties of the materials but also on the properties of the laser radiation.
 The increased number of parameters describing multi-orbital physics enriches the possibilities for Floquet engineering in such systems.   Applications to the orthorhombic titanates are studied in \cite{liu2018floquet}, where it is shown there by realistic calculations that the sign of the static exchange interaction can be flipped in the Floquet regime if one uses radiation with certain values for the frequency and electric field strength.

We have further argued that if the frequency of the radiation (and its multiples) is kept away from the Hubbard interaction scale, collectively shown as $U$, the heating due to creation of doublon holon pairs can be avoided and an effective spin or spin-orbital (in the multi-orbital case) Hamiltonian can capture the physics of the system. This is due to the finite bandwidth of the excitations, i.e.~doublon-holon pairs, which we have shown is of order $\sqrt{z-1} t_h$. This gives a criterion for how distant one needs the laser frequency to be from the interaction energy in the Hamiltonian describing the material. We have also given relations for the doublon holon creation rates, which are derived by taking the finite bandwidth of excitations into account. Using these relations, we have also studied the rate of creation of DH pairs in very small frequency regime and derived a zero-frequency limit of absorption which can be related to the breakdown of the Mott insulator in presence of a static electric field.  The maximum rate occurs when the electric field energy between neighboring sites is comparable to the Hubbard interaction energy scale.  This may be observable experimentally.  A potentially interesting subject for future theoretical study is the effects of {\em selective} doublon-holon generation with specific quantum numbers, i.e. preferential generation of singlets or triplets, forming a gas of excitations with controllable internal degrees of freedom.   Our equations for the generation rates of these excitations provide a starting point for such a study.

\section*{ACKNOWLEDGEMENTS}
K.H. thanks D. Else and M. Kolodrubetz for fruitful discussions. This research was supported by the NSF materials theory program through grant DMR1818533 (LB, KH) and the Army Research Office MURI grant ARO W911NF-16-1-0361, Floquet engineering and metastable states (JL).

\bibliography{yto}

\onecolumngrid
\appendix
\section{Time averaging the virtual hoppings}\label{app:time_averaging}
In this appendix, we justify the time averaging of virtual hoppings  in the time dependent Schr\"odinger equations \eqref{eq:truncated_eom} and \eqref{eq:truncated_eom_multi_t_s} of the main text.

We first consider the single-orbital case. Consider the left hand side of the second line of equation \eqref{eq:truncated_eom} written in a different way:
\begin{equation}\label{eq:psi1_eom_th_ave}
	i \left( \partial_t + i U + i \tilde{T}_t\right)|\Psi_{1}\rangle_t =   \ T_t |\Psi_0\rangle_t.
\end{equation}
The unitary evolution operator $S_t = \hat{P}_1 S_t \hat{P}_1$, defined as satisfying an equation analogous to the above:
\begin{equation}\label{eq:defin_S}
	i \left( \partial_t + i U + i \tilde{T}_t\right) S_t = 0,
\end{equation}
can be useful, in the sense that if one finds $S_t$, the solution to the equation \eqref{eq:defin_S}, with initial condition $S_{t=0} = 1$ one can write \eqref{eq:psi1_eom_th_ave} as:
\begin{equation}\label{eq:1}
S_t \	i  \partial_t   \left(S_t^{-1}|\Psi_{1}\rangle_t \right) =   \ T_t |\Psi_0\rangle_t,
\end{equation}
where $S^{-1}$ only acts on states in the subspace invariant under $\hat{P}_1$.
The equation of motion of $S_t$ can be written as:
\begin{equation}\label{eq:interaction_picutre_eom_S}
	 \ i  \partial_t   \left(e^{iUt}\ S_t\right) = e^{iUt}\ \tilde{T}_t  \ e^{-iUt} \ \left(e^{iUt}\ S_t \right).
\end{equation}
This is similar to an interaction picture time evolution. Noting that the exponentials commute with $\tilde{T}_t$, one notices that the above equation is of Floquet type and thus the corresponding effective Hamiltonian to leading order is obtained by time averaging\cite{eckardt2015high,bukov-aip-15}. This fact can also be seen by the following manipulations.   Using the Fourier series for the hopping operator $\tilde{T}_t = \sum_{n} e^{in\omega t} \tilde{T}_n$, eq. \eqref{eq:interaction_picutre_eom_S} reads:
\begin{equation}
	 \ i  \partial_t   \left(e^{iUt}\ S_t\right) = \sum_{n} e^{in\omega t} \tilde{T}_n \ \left(e^{iUt}\ S_t\right).
\end{equation}
Noting $\tilde{T}_0 = \bar{T}$, one has:
\begin{equation}
	  i \, e^{-i \bar{T}t} \,  \partial_t   \left(e^{i(U + \bar{T})t}\ S_t\right) = \sum_{n \neq 0} e^{in\omega t} \tilde{T}_n \ \left(e^{iUt}\ S_t\right).
\end{equation}
Moving the factor $e^{-i \bar{T}t}$ to the right hand side, one can integrate the above equation and do integration by parts on the right hand side, keeping in mind that the derivative of $\left(e^{iUt}\ S_t\right)$ is of order $t_h$ (this is similar to what was done in the main text):
\begin{equation}
	\begin{aligned}
	i \left[ \left.\left(e^{i(U + \bar{T})t'}\ S_{t'}\right)\right|_{t'=t} -  S_{t'=0} \right] &= \sum_{n \neq 0} \frac{1}{i(\bar{T}+n\omega)} \Bigg\{ \left[ \left.e^{i(\bar{T} + n\omega)t'} \tilde{T}_n \left(e^{i U t'}\ S_{t'}\right)\right|_{t'=t} - \tilde{T}_n   S_{t'=0} \right]\\
	&\qquad \qquad \qquad \qquad \qquad -\int_0^t dt' \; e^{i(\bar{T} + n\omega)t'} \tilde{T}_n \partial_{t'} \left(e^{i U t'}\ S_{t'}\right)  \Bigg\} \\
	 &= \mathcal{O}\left( \frac{t_h}{\omega} \right).
	\end{aligned}
\end{equation}
The first line on the right hand side is of first order and the second line of second order and thus the right hand side is first order over all.

Finally, the solution can be obtained to leading order:
\begin{equation}
	S_t = e^{-i\left( U + \bar{T} \right) t} + \mathcal{O}\left( \frac{t_h}{\omega} \right),
\end{equation}
and plugging this back into \eqref{eq:1}, one obtains the desired result.

One can generalize this to the multi-orbital case also, noting the fact that the multi-orbital doublon can be in a singlet or triplet state and so the following decompositions should be considered:
\begin{equation}
\hat{P}_1 = \hat{P}_1^\mathrm{s} + \hat{P}_1^\mathrm{t}, \qquad \hat{U} = U \hat{P}_1^\mathrm{s} + (U - 2J) \hat{P}_1^\mathrm{t},
\end{equation}
\begin{equation}
\tilde{T}_t = \tilde{T}_t^{\mathrm{ss}} + \tilde{T}_t^{\mathrm{st}} + \tilde{T}_t^{\mathrm{ts}} + \tilde{T}_t^{\mathrm{tt}},
\end{equation}
with $\tilde{T}_t^{\mathrm{ab}} = \hat{P}_1^a \tilde{T}_t \hat{P}_1^b$.
The analogue of eq. \eqref{eq:interaction_picutre_eom_S}, can be written as:
\begin{equation}
	 \begin{aligned}
	 	\ i  \partial_t   \left(e^{i\hat{U}t}\ \hat{S}_t\right) &= e^{i\hat{U}t}\ \tilde{T}_t  \ e^{-i\hat{U}t} \ \left(e^{i\hat{U}t}\ \hat{S}_t \right) \\
	 	&= \left[  \tilde{T}_t^{\mathrm{ss}} + \tilde{T}_t^{\mathrm{tt}} + \tilde{T}_t^{\mathrm{ts}}\,  e^{i2Jt} + \tilde{T}_t^{\mathrm{st}}\,  e^{-i2Jt} \right]  \left(e^{i\hat{U}t}\ \hat{S}_t \right).
	 \end{aligned}
\end{equation}
A similar argument like the one carried out for the case of a single orbital can be applied here also, except that when $2J$ and $n\omega$ are not close to each other (compared with $t_h$), $ \tilde{T}_t^{\mathrm{ts}}\,  e^{i2Jt}$ and $\tilde{T}_t^{\mathrm{st}}\,  e^{-i2Jt}$ do not have constant terms. Thus the final form for $\hat{S}_t$ reads:
\begin{equation}
	\hat{S}_t = e^{-i\left( \hat{U} + \bar{T}^{\mathrm{ss}} + \bar{T}^{\mathrm{tt}} \right) t} + \mathcal{O}\left( \frac{t_h}{\omega_0} \right),
\end{equation}
where $t_h$ in $\mathcal{O}$ show the typical hopping parameter and $\omega_0$ in the denominator stands for either of $\omega$ and $J$.

\section{Overview of the Retraceable Path approximation}\label{app:RP}
In this appendix we present a short discussion of the Retraceable Path (RP) approximation of Brinkman and Rice \cite{brinkman-rice-prb70}. With the notation of the main text, one can write the Green's function of a single hole in a single band hubbard model as:
\begin{equation}
	\begin{aligned}
		g_{\text{h}}(E) &= \sum_\sigma \langle \Psi_0 \lvert   c^\dagger_{j\sigma} \ \frac{1}{E + \bar{T}_{\text{h}}}  \ c_{j\sigma} \rvert \Psi_0 \rangle\\
			&=  \frac{1}{E} \langle 0 \lvert c_{j\sigma_j}^\dagger  \ \left[\, 1 + \left( - \frac{\bar{T}_{\text{h}}}{E} \right) + \left( - \frac{\bar{T}_{\text{h}}}{E} \right)^2 + \left( - \frac{\bar{T}_{\text{h}}}{E} \right)^3 + \ldots  \, \right] \  c_{j\sigma_j} \rvert 0 \rangle.
	\end{aligned}
\end{equation}
Note that $\rvert \Psi_0 \rangle$ has one electron per site and thus the state $c_{j\sigma_j} \rvert \Psi_0 \rangle$ has a hole at site $j$, with $\sigma_j$ showing the spin at site $j$ in state $\rvert \Psi_0 \rangle$. The series on the right hand side shows that one should consider all the possible paths including arbitrary number of holon hoppings that connect the state with a hole to itself. Moreover, the final spin configuration should be the same as the initial. The RP approximation amounts to considering only paths that start at $j$ and terminate at the same point, with the constraint that the hopping holon should exactly retrace its forward going path in its way back to the original location. With this constraint, every spin reordering that is done in the forward going path is corrected when the particle is getting back. What one is missing here is the contribution by the paths that are closed loops and correct all the spin reorderings in some way.

We use the following ansatz for the one holon Green funcion, with the introduction of a self energy:
\begin{equation}
	g_{\text{h}}(E) = \frac{1}{E \left[ 1-\Sigma(E)\right]}.
\end{equation} 
Since we are considering paths with any number of hoppings, at each step of a path for the remainder of the path, one should consider all the paths that start at the given point and come back to the same position, except for the one going backward. This is very similar to what we are trying to compute, and thus in order to perform an infinite summation over the retraceable paths, one introduces a summation of all forward going paths at a specific step of the path. At a given step of the process, since the paths can just go forward there are $(z-1)$ choices for direction of the next step, with $z$ the coordination number. The following equation for $\Sigma^A$, will result in a self-consistent summation over forward going paths with arbitrary lengths, something that is present as the future of every step (other than the first) in a retraceable path:
\begin{equation}
	\Sigma^A(E) = \frac{(z-1) t_h^2}{E^2 \left[1-\Sigma^A(E) \right]},
\end{equation}
which has the solution:
\begin{equation}\label{eq:sol_sigma_A}
	\Sigma^A(E) = \frac12 \left[ 1 \pm \sqrt{1 - 4 (z-1) \frac{t_h^2}{E^2}}  \right].
\end{equation}
Where the hopping parameter in $\bar{T}_h$ is shown as $t_h$. The self-energy can be written in terms of the sum of all the forward going paths as:
\begin{equation}
	\Sigma(E) = \frac{z}{z - 1} \Sigma^A(E),
\end{equation}
because at the first step of each path there are $z$ choices of direction rather than $z-1$ choices for the holon. This finally results in the following form for the Green's function:
\begin{equation}
	g_{\text{h}} (E) =\frac{2(z-1)}{E \left[(z-2)+z\sqrt{1-4(z-1) t_h^2/E^2} \right] }\;,
\end{equation}
The solution with a minus sign in equation \eqref{eq:sol_sigma_A}, is chosen so that the above Green's function behaves as $\frac{1}{E}$ as $E \to \infty$.

One can also consider the case of anisotropic hoppings which results in solving more self-consisten equations. We will work with hopping parameters $t_\mu$ that are different for different directions. One can further introduce a summation over all the forward going paths in different directions, denoted as $\Sigma^A_\mu (E)$. Which should satisfy the following self-consistent equations:
\begin{equation}\label{eq:self_consistent_anisotropic}
	\Sigma^A_\mu(E) = \frac{t_\mu^2/E^2}{1 - \Sigma^A_\mu (E) - \sum_{\mu' \neq \mu} \Sigma^A_{\mu'}(E)}, \qquad \mu = 1 , \ldots , d.
\end{equation}
And the total self energy used for obtaining the Green's function can be written in terms of the $\Sigma_\mu^A$ as follows:
\begin{equation}
\Sigma(E) = 2 \sum_{\mu} \Sigma_\mu^A(E).
\end{equation}

A first approximation for finding the solution to the above self consistent equations \eqref{eq:self_consistent_anisotropic} would be to use the average hopping $\sum_\mu t_\mu$ for every $t_\mu$; one can make this approximation better by iterating the solution obtained this way in equations \eqref{eq:self_consistent_anisotropic}.

\section{The frequency integral of the DH Green's function}\label{app:convolution}

In this appendix, we show how the convolution integral in equation \eqref{eq:g_dh_def} can be done in order the get the functional form shown in figure \ref{fig:gdh_calc}.

\begin{figure}[!t]
\centering
\subfloat[Definition of square root in the complex $E$ plane.\label{fig:def_sqrt}]{%
  \includegraphics[width=.4\linewidth]{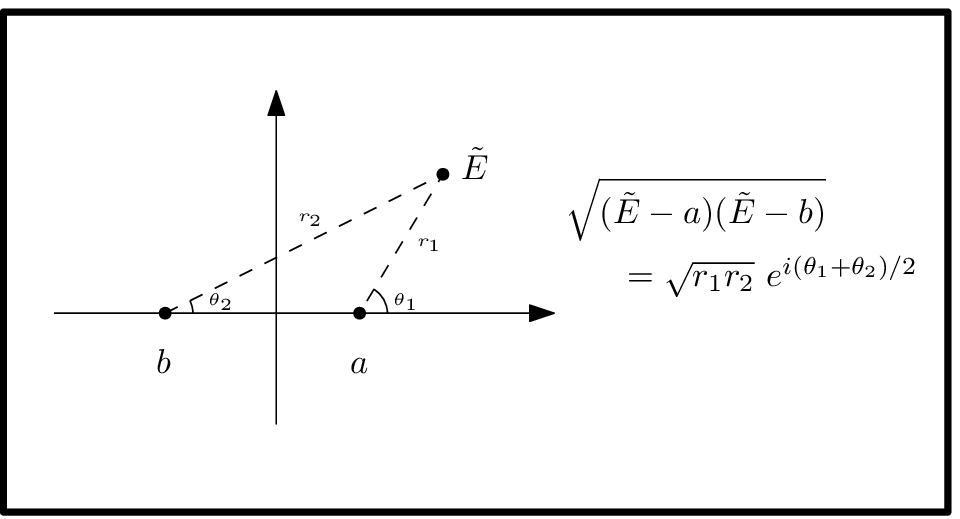}%
}\qquad
\subfloat[\label{fig:branches}The contour for computing $g_{\text{RP}}^{\text{dh}}$.]{%
  \includegraphics[width=.4\linewidth]{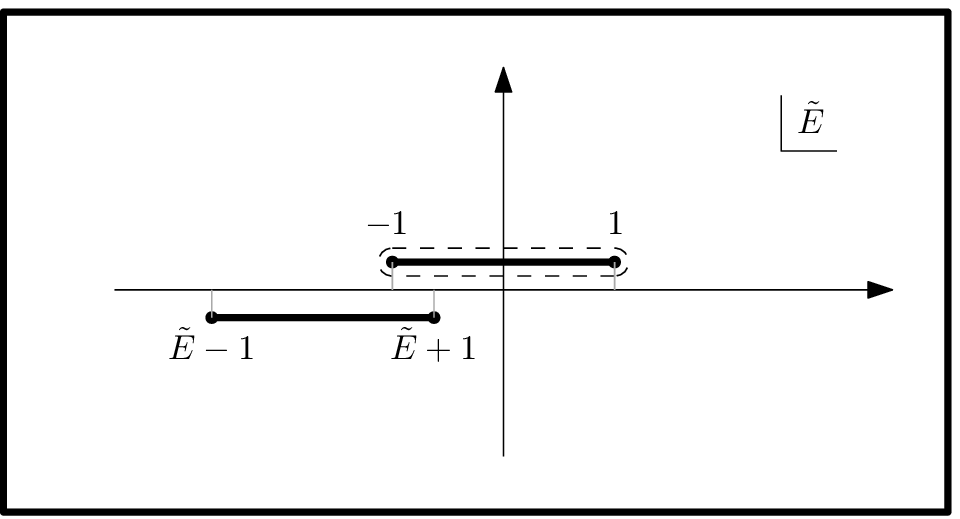}%
} \caption{Appendix C figures. }
\end{figure}

The holon and doublon Green's functions in the RP approxiamtion can be written in the following form as well:
\begin{equation}
	g_{\text{h}}(E ) = g_{\text{d}}(E ) = \frac{2 \ (z-1)}{E(z-2) + z \sqrt{ E^2 - 4(z-1)\tilde{t}_h^2 } },
\end{equation}
with the definition of the square root function in the denominator, in the complex $E$ plane, presented in Figure \ref{fig:def_sqrt}. It is easy to check that with this definition and the signs above, the two Green's functions fall off like $\frac{1}{E}$ for large $E$. Note that we define the dimensionless frequencies as $\tilde{E} = \frac{E}{2\tilde{t}_h \sqrt{z-1}}$.

In order to find $g_{\text{dh}}$, one should do the frequency integral in equation \eqref{eq:g_dh_def}, with the integrand having two branch cuts as shown in Figure \ref{fig:branches}. Using Cauchy's theorem the contour $(-\infty,\infty)$ can be deformed into a contour that turns around the upper branch cut, and the integral can be done for this contour.  In four different ranges for $E$ the integral over this contour is computed in the following:

\begin{itemize}
	\item Case $i$) $\quad\tilde{E} > 2$

	\begin{equation}
		\left( \frac{\pi}{2} \ \tilde{t}_h \ \sqrt{z-1} \right)   \  g_{\text{dh}}(\tilde{E}) = (z-1)^2 z \int_{-1}^1 d\tilde{\Omega}  \ \frac{\left(  1 - \tilde{\Omega}^2  \right)^{1/2}}{\tilde{\Omega}^2 (- 4z + 4) + z^2 } \ \frac{1}{(\tilde{E} - \tilde{\Omega})(z-2) + z \left(  (- \tilde{E} + \tilde{\Omega})^2 - 1  \right)^{1/2}}.
	\end{equation}
		
	\item Case $ii$) $\quad \tilde{E} < -2$
	
	\begin{equation}
		\left( \frac{\pi}{2} \ \tilde{t}_h \ \sqrt{z-1} \right)   \  g_{\text{dh}}(\tilde{E}) = (z-1)^2 z \int_{-1}^1 d\tilde{\Omega}  \ \frac{\left(  1 - \tilde{\Omega}^2\right)^{1/2}}{\tilde{\Omega}^2 (- 4z + 4) + z^2 } \ \frac{1}{(\tilde{E} - \tilde{\Omega})(z-2) - z  \left(  (- \tilde{E} + \tilde{\Omega})^2 - 1  \right)^{1/2}}.
	\end{equation}

	\item Case $iii$) $\quad 0 < \tilde{E} < 2$
	
	\begin{equation}
		\begin{aligned}
			\left( \frac{\pi}{2} \ \tilde{t}_h \ \sqrt{z-1} \right)   \  g_{\text{dh}}(\tilde{E})
				&= (z-1)^2 z \int_{-1}^{\tilde{E} - 1} d\tilde{\Omega}  \ \frac{\left(  1 - \tilde{\Omega}^2\right)^{1/2}}{\tilde{\Omega}^2 (- 4z + 4) + z^2 } \ \frac{1}{(\tilde{E} - \tilde{\Omega})(z-2) + z  \left(  (- \tilde{E} + \tilde{\Omega})^2 - 1  \right)^{1/2} }\\
				&+ (z-1)^2 z \int^{1}_{\tilde{E} - 1} d\tilde{\Omega}  \ \frac{\left(  1 - \tilde{\Omega}^2\right)^{1/2}}{\tilde{\Omega}^2 (- 4z + 4) + z^2 } \ \frac{1}{(\tilde{E} - \tilde{\Omega})(z-2) - i z \left(   1 - (- \tilde{E} + \tilde{\Omega})^2  \right)^{1/2} }.
		\end{aligned}
	\end{equation}
	
	\item Case $iv$)   $\quad -2 < \tilde{E} < 0$

	\begin{equation}
		\begin{aligned}
			\left( \frac{\pi}{2} \ \tilde{t}_h \ \sqrt{z-1} \right)   \  g_{\text{dh}}(\tilde{E}) 
			&= (z-1)^2 z \int_{\tilde{E}+1}^{ 1} d\tilde{\Omega}  \ \frac{\left(  1 - \tilde{\Omega}^2\right)^{1/2}}{\tilde{\Omega}^2 (- 4z + 4) + z^2 } \ \frac{1}{(\tilde{E} - \tilde{\Omega})(z-2) - z  \left(  (- \tilde{E} + \tilde{\Omega})^2 - 1  \right)^{1/2} }\\
			&+ (z-1)^2 z \int^{\tilde{E}+1}_{ - 1} d\tilde{\Omega}  \ \frac{\left(  1 - \tilde{\Omega}^2\right)^{1/2}}{\tilde{\Omega}^2 (- 4z + 4) + z^2 } \ \frac{1}{(\tilde{E} - \tilde{\Omega})(z-2) - i z\left(   1 - (- \tilde{E} + \tilde{\Omega})^2  \right)^{1/2} }.
		\end{aligned}
	\end{equation}

\end{itemize}

$g_{\text{dh}}$ only has nonzero imaginary part in the case $iii$ and $iv$. A plot of $g_{\text{dh}}$, obtained above can be found in Figure \ref{fig:gdh_calc}. $z=6$ is taken for this plot.

\section{Zero-frequency limit of DH pairs creation rate}\label{app:asymptotic}
\begin{figure}[!b]
\centering
\subfloat[\label{fig:zeta}\footnotesize The function $\zeta(z)$ defined in \eqref{eq:zeta}. It is positive when $z < 1$ and negative when $z > 1$ and vanishes at $z=0$.]{%
  \includegraphics[width=.4\linewidth]{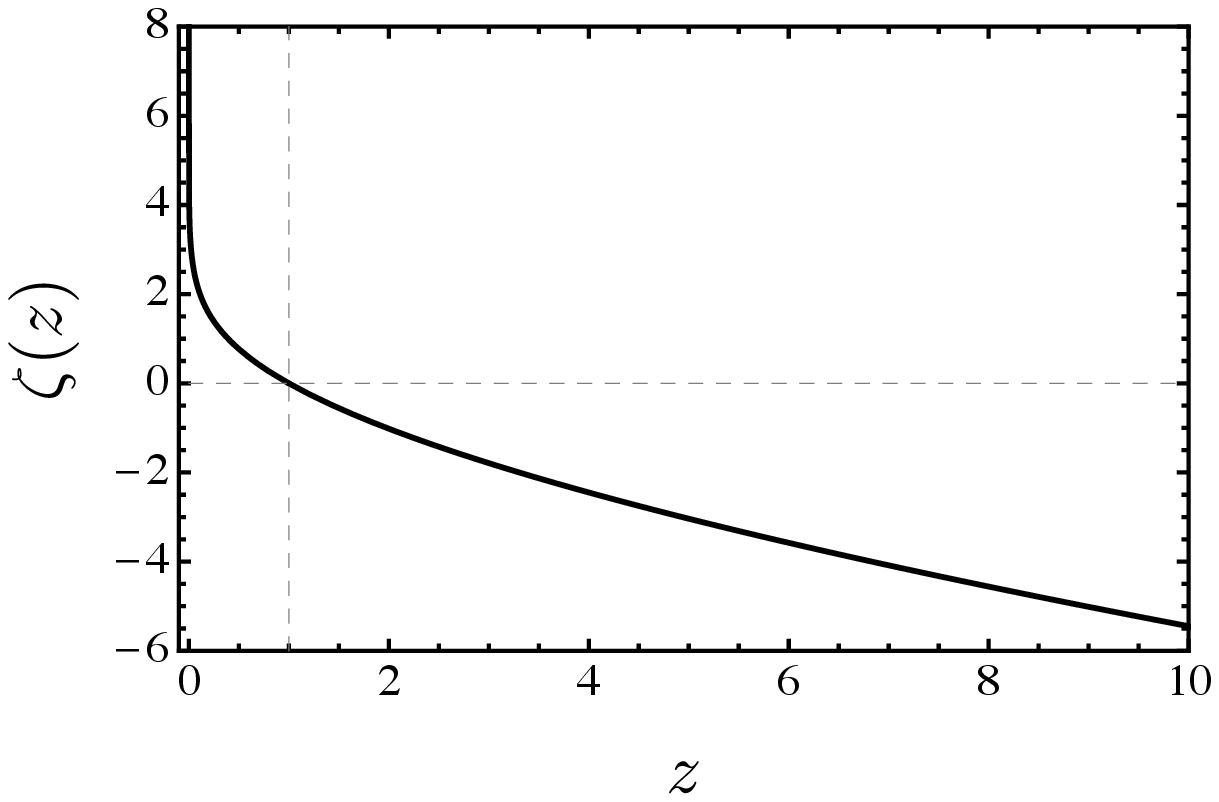}%
}\qquad
\subfloat[\label{fig:f}\footnotesize The function $f(x) = \sqrt{1-x^2}-x \, \cos ^{-1}(x)$ is a monotonic function for $ 0 < x < 1$.]{%
  \includegraphics[width=.4\linewidth]{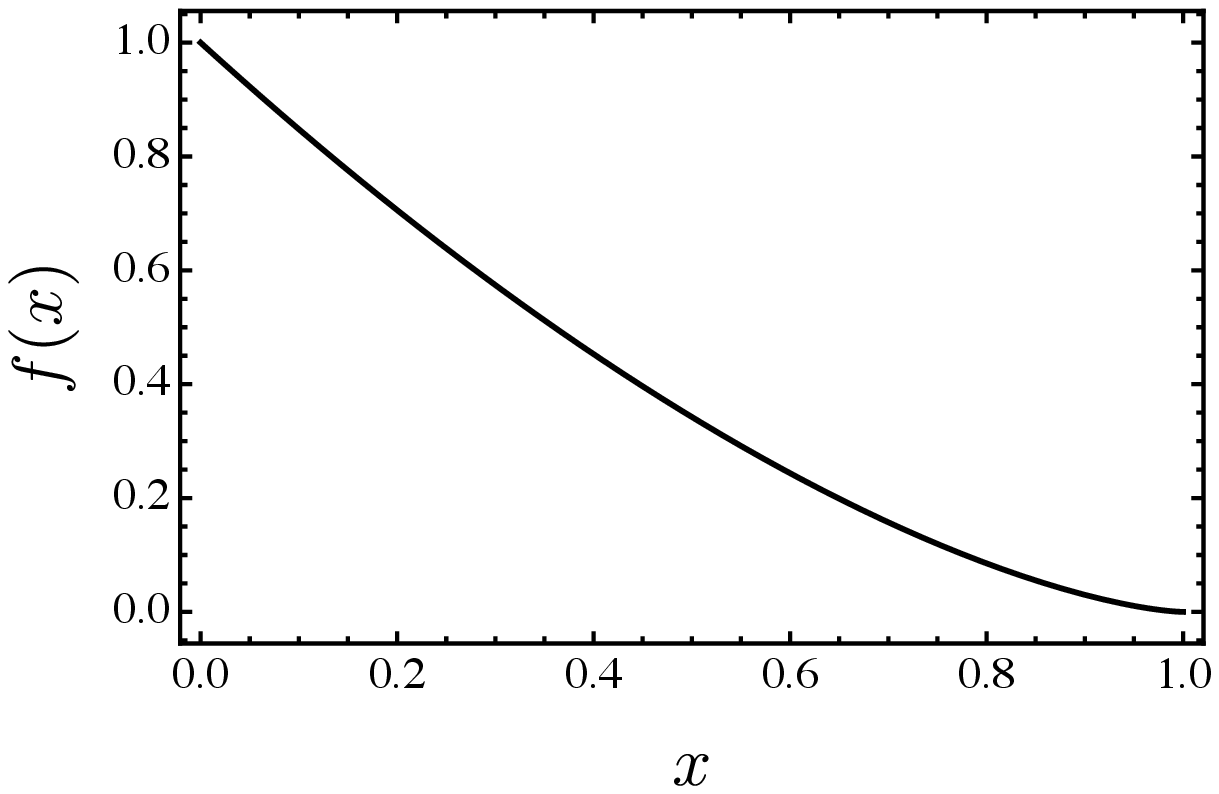}%
}\caption{The two auxiliary functions used in this appendix.}
\end{figure}

In this appendix, we present a derivation of the zero-frequency limit of the DH pair creation rate, and derive the result \eqref{eq:final_form_S}. We will take the frequency $\omega$ to be very small. Our starting point is the sum:
\begin{equation}\label{eq:sum_bessel_append}
	\mathcal{S} = \sum_{n=-\infty}^\infty \mathcal{J}_n^2(\Phi/\omega)\, \mathrm{Im} g_{\text{dh}}(U - n \omega) = \sum_{n \sim U/\omega - 2 /\omega}^{U/\omega + 2 /\omega}  \mathcal{J}_n^2(\Phi/\omega) \, \mathrm{Im} g_{\text{dh}}(U - n \omega).
\end{equation}
Since $n$ is a very large number in all of the terms in the above sum, one is able to use the following high order Bessel function asymptotic form \cite{NIST:DLMF,temme1997numerical}:
\begin{equation}\label{eq:bessel_asymptotic}
	\mathcal{J}_\nu(x) \sim \left(\frac{4\, \zeta \left(\frac{x}{\nu }\right)}{1-\left(\frac{x}{\nu }\right)^2}\right)^{1/4}  \frac{\text{Ai}\left(\nu^{2/3} \, \zeta \left(\frac{x}{\nu }\right)\right)}{\nu^{1/3} },
\end{equation}
which holds for large and positive $\nu$ and positive $x$, and $\text{Ai}$ is the Airy function. The function $\zeta(z)$ is defined as:
\begin{equation}\label{eq:zeta}
	\zeta(z) = \begin{cases} 
      \left[\frac{3}{2} \left(\log \left(\frac{\sqrt{1-z^2}+1}{z}\right)-\sqrt{1-z^2}\right)\right]^{2/3} & z\leq 0, \\
      -\left[\frac{3}{2} \left\{\sqrt{z^2-1}-\cos ^{-1}\left(\frac{1}{z}\right)\right\}\right]^{2/3} & z>1.
          \end{cases}
\end{equation}
$\zeta(z)$ is positive if $z<1$ and negative if $z>1$. The function $\zeta(z)$ is depicted in figure \ref{fig:zeta}, but we will not need its exact form finally.
Plugging the asymptotic form back into equation \eqref{eq:sum_bessel} and converting the sum over $n$ into an integral:
\begin{equation}\label{eq:s_integral}
	\begin{aligned}
		\mathcal{S} &= \int_{U/\omega - 2/\omega}^{U/\omega + 2/\omega} d\nu \; \left(\frac{4\, \zeta \left(\frac{\Phi}{\omega\nu }\right)}{1-\left(\frac{\Phi}{\omega\nu }\right)^2}\right)^{1/2}  \frac{\text{Ai}^2\left(\nu^{2/3} \, \zeta \left(\frac{\Phi}{\omega\nu }\right)\right)}{\nu^{2/3} } \  \mathrm{Im} g_{\text{dh}}(U - \nu \omega)\\
		& = \int_{U - 2}^{U + 2} d\bar{\nu} \; \left(\frac{4\, \zeta \left(\frac{\Phi}{\bar{\nu} }\right)}{1-\left(\frac{\Phi}{\bar{\nu} }\right)^2}\right)^{1/2}  \frac{1}{\bar{\nu}^{2/3} } \  \mathrm{Im} g_{\text{dh}}(U - \bar{\nu}) \left[ \frac{1}{\omega^{1/3}} \, \text{Ai}^2\left(\bar{\nu}^{2/3}\, \omega^{-2/3} \, \zeta \left(\frac{\Phi}{\bar{\nu} }\right)\right) \right]
	\end{aligned}
\end{equation}
where the substitution $\bar{\nu} = \omega \nu$ is done in the second line. Only the expression in the $[.]$ is $\omega$-dependent, and thus it is the term that should be studied in the limit $\omega \to 0$. The following asymptotic forms for the Airy function with a large argument $\left| x \right| \gg 1$ will be used:
\begin{equation}
	\text{Ai}(x) \sim \begin{cases}
	\frac{1}{2 \sqrt{\pi }}\, x^{-1/4} \, e^{- \frac{2}{3}  x^{3/2}} ,    &  x > 0,\\
	\frac{1}{\sqrt{\pi }} \, (-x)^{-1/4} \, \sin \left(\frac{2}{3} (-x)^{3/2}+\frac{\pi }{4}\right) ,    & x < 0.
\end{cases}
\end{equation}

 The expression containing $\omega$ in equation \eqref{eq:s_integral} can now be studied in the limit $\omega \to 0$ with the above asymptotic forms; for positive values of $\zeta(\Phi/\bar{\nu})$ the Airy function decays exponentially in $1/\omega$, and thus one should exclude the regions corresponding to $\zeta (\Phi/\bar{\nu}) > 0$ in the integral \eqref{eq:s_integral}, while for negative values of $\zeta(\Phi/\bar{\nu})$ one has:
\begin{equation}
	\begin{aligned}
		\frac{1}{\omega^{1/3}} \, \text{Ai}^2\left(\bar{\nu}^{2/3}\, \omega^{-2/3} \, \zeta \left(\frac{\Phi}{\bar{\nu} }\right)\right) &\sim \frac{1}{\omega^{1/3}} \left[ \frac{\omega^{1/3}}{\pi \, \bar{\nu}^{1/3}} \frac{1}{\left(-\zeta(\Phi/\bar{\nu}) \right)^{1/2}}\right] \\
		& \qquad \qquad \times \sin^2\left\{\frac{2}{3} \left(-\bar{\nu}^{2/3}\, \omega^{-2/3} \, \zeta \left(\Phi / \bar{\nu} \right)\right)^{3/2}+\frac{\pi }{4} \right\}.
	\end{aligned}
\end{equation}
The function $\zeta$ is negative when its argument is larger than $1$, which means $\Phi>\bar{\nu}$. Furthermore, the argument of the $\sin^2$ function in the above asymptotic form can be rewritten using the definition of $\zeta$ as: 
\begin{equation}
	\begin{aligned}
	\frac{2}{3} \left(-\bar{\nu}^{2/3}\, \omega^{-2/3} \, \zeta \left(\Phi / \bar{\nu} \right)\right)^{3/2} 
		&= \frac{\bar{\nu}}{\omega} \, \frac{2}{3} \left(- \zeta \left(\Phi / \bar{\nu} \right)\right)^{3/2} \\
		&= \frac{\bar{\nu}}{\omega} \, \frac{2}{3} \left(\left[\frac{3}{2} \left\{\sqrt{\left(\Phi / \bar{\nu} \right)^2-1}-\cos ^{-1}\left(\frac{1}{\left(\Phi / \bar{\nu} \right)}\right)\right\}\right]^{2/3}\right)^{3/2} \\
		&= \frac{\Phi}{\omega} \left[ \sqrt{1 - \left( \frac{\bar{\nu}}{ \Phi} \right)^2} - \left( \frac{\bar{\nu}}{ \Phi} \right)\, \cos ^{-1}\left( \frac{\bar{\nu}}{ \Phi} \right) \right]\\
		& = \frac{\Phi}{\omega} \	f \left( \frac{\bar{\nu}}{ \Phi} \right) .
	\end{aligned}
\end{equation}
The function $f$ is defined as $f(x) = \sqrt{1-x^2}-x \cos ^{-1}(x)$. It is a monotonic function of its argument (figure \ref{fig:f}), when $0<x<1$, because $df/dx = - \cos^{-1}(x)$. We will use this point in what follows. Plugging the asymptotic form back into equation \eqref{eq:s_integral}, one has:
\begin{equation}
	\begin{aligned}
		\mathcal{S} & = \int_{U - 2}^{\Phi} d\bar{\nu} \; \left(\frac{4\, \zeta \left(\frac{\Phi}{\bar{\nu} }\right)}{1-\left(\frac{\Phi}{\bar{\nu} }\right)^2}\right)^{1/2}  \frac{1}{\bar{\nu}^{2/3} } \  \mathrm{Im} g_{\text{dh}}(U - \bar{\nu}) \left[ \frac{1}{\pi \bar{\nu}^{1/3} \, \left(-\zeta(\Phi/\bar{\nu}) \right)^{1/2}} \right] \\
		 &\qquad\qquad \times \quad \sin^2\left[\frac{\Phi}{\omega} \	f \left( \frac{\bar{\nu}}{ \Phi} \right)+\frac{\pi }{4} \right]\\
		&= \frac{2}{\pi} \int_{U - 2}^{\Phi} d\bar{\nu} \; \left(\frac{1}{\left(\frac{\Phi}{\bar{\nu} }\right)^2 - 1}\right)^{1/2}  \frac{1}{\bar{\nu} } \  \mathrm{Im} g_{\text{dh}}(U - \bar{\nu}) \ \sin^2\left[\frac{\Phi}{\omega} \	f \left( \frac{\bar{\nu}}{ \Phi} \right)+\frac{\pi }{4} \right].
	\end{aligned}
\end{equation}
Since the function $f$ does not have vanishing derivative in the domain of integration, and we are interested in the $\omega \to 0$ limit, one can argue that the function $\sin^2$ oscillates very rapidly and thus can be substituted by its average value $\frac{1}{2}$. One finally can write $\mathcal{S}$ as:
\begin{equation}\label{eq:s_final}
	\mathcal{S} = \frac{1}{\pi} \int_{U - 2}^{\Phi} d\bar{\nu} \; \left(\frac{1}{\left(\frac{\Phi}{\bar{\nu} }\right)^2 - 1}\right)^{1/2}  \frac{1}{\bar{\nu} } \  \mathrm{Im} g_{\text{dh}}(U - \bar{\nu}).
\end{equation}

\end{document}